\documentclass[12pt]{article}
\usepackage[reqno]{amsmath}
\usepackage{epsfig}
\usepackage{array}
\usepackage{float}
\usepackage{lscape,graphicx}
\usepackage{graphics}
\usepackage{amssymb}
\usepackage{color}
\usepackage{booktabs}
\usepackage{amsfonts}
\usepackage{hyperref}
\hypersetup{citecolor=blue}
\usepackage{colordvi}

\def\ltap{\ \raisebox{-.4ex}{\rlap{$\sim$}} \raisebox{.4ex}{$<$}\ }
\def\gtap{\ \raisebox{-.4ex}{\rlap{$\sim$}} \raisebox{.4ex}{$>$}\ }

\newcommand{\Dmq}{\ensuremath{\Delta m^2}}

\newcommand{\be}{\begin{equation}}
\newcommand{\ee}{\end{equation}}

\newcommand{\bea}{\begin{equation}\begin{array}{c}}
\newcommand{\eea}{\end{array}\end{equation}}
\newcommand{\ea}{\end{array}}

\newcommand{\beq}{\begin{equation}}
\newcommand{\eeq}{\end{equation}}
\newcommand{\bad}{\begin{array}{ccc}}

\newcommand{\ba}{\begin{array}{c}}
\begin{document}
\hfill{{\small Ref. SISSA 64/2014/FISI}}

\begin{center}
\mathversion{bold}
{\bf{\Large Radiative Emission of Neutrino Pairs in Atoms and
Light Sterile Neutrinos}}
\mathversion{normal}

\vspace{0.4cm}
D. N. Dinh$\mbox{}^{a)}$,
S. T. Petcov$\mbox{}^{b,c)}$~
\footnote{Also at: Institute of Nuclear Research and
Nuclear Energy, Bulgarian Academy of Sciences, 1784 Sofia, Bulgaria}

\vspace{0.2cm}
$\mbox{}^{a)}${\em Institute of Physics, Vietnam Academy of Science and Technology, \\ 10 Dao Tan, Hanoi, Vietnam.\\
}

\vspace{0.1cm}
$\mbox{}^{b)}${\em  SISSA and INFN-Sezione di Trieste, \\
Via Bonomea 265, 34136 Trieste, Italy.\\}

\vspace{0.1cm}
$\mbox{}^{c)}${\em Kavli IPMU, University of Tokyo (WPI), Tokyo, Japan.\\
}
\end{center}

\begin{abstract}
 The process of Radiative Emission of Neutrino Pair
(RENP) in atoms is sensitive to the absolute neutrino mass scale,
the type of spectrum neutrino masses obey and  the nature -
Dirac or Majorana - of massive neutrinos.
We analise the possibility to test
the hypothesis of exsitence of
neutrinos with masses at the eV scale
coupled to the electron in the weak charged
lepton current in an RENP experiment.
The presence of eV scale neutrinos in the
neutrino mixing is associated with the
existence of sterile neutrinos
which mix with the active flavour
neutrinos. At present there are a number
of hints for active-sterile neutrino
oscillations driven by
$\Delta m^2 \sim 1~{\rm eV^2}$.
We perform a detailed analysis of the RENP
phenomenology within the
``3 + 1'' scheme with one sterile neutrino.
\end{abstract}

\section{Introduction}

  All compelling neutrino oscillation data can be described
within the reference 3-flavour neutrino mixing scheme
with 3 light neutrinos $\nu_j$ having masses $m_j$ not
exceeding approximately 1 eV, $m_j \ltap 1$ eV, $j=1,2,3$
(see, e.g., \cite{PDG2012}). These data allowed to determine
the parameters which drive the observed solar,
atmospheric, reactor and accelerator flavour
neutrino oscillations - the three neutrino mixing angles
of the standard parametrisation of the Pontecorvo, Maki,
Nakagawa and Sakata (PMNS) neutrino mixing matrix,
$\theta_{12}$, $\theta_{23}$ and $\theta_{13}$, and the two
neutrino mass squared differences $\Delta m^2_{21}$ and
$\Delta m^2_{31}$ (or $\Delta m^2_{32}$) - with a relatively
high precision \cite{Capozzi:2013csa,GonzalezGarcia:2012sz}.

  Although the mixing of the 3 flavour neutrino states
has been experimentally well established,
implying the existence of 3 light neutrinos $\nu_j$
having masses $m_j \ltap 1$ eV,
there have been possible hints for the presence in the mixing
of one or more additional neutrino states with masses at the eV scale.
If these states exist, they must be related to the existence of
one or more sterile neutrinos (sterile neutrino fields)
which mix with the active flavour neutrinos
(active flavour neutrino fields).
The hints in question have been obtained:
i) in the LSND $\bar{\nu}_{\mu} \rightarrow \bar{\nu}_e$
appearance experiment \cite{Aguilar01LSND},
in which a significant
excess of events over the background is claimed
to have been observed,
ii) from the analysis of the
$\bar{\nu}_{\mu} \rightarrow \bar{\nu}_e$
and $\nu_{\mu} \rightarrow \nu_e$ appearance
data of the MiniBooNE experiment \cite{MiniBooNE10,MiniBooNE13},
iii) from the re-analyses of the
short baseline (SBL) reactor neutrino oscillation
data using newly calculated fluxes of reactor
$\bar{\nu}_e$ \cite{Mention11,mueller11},
which show a possible ``disappearance''  of
the reactor $\bar{\nu}_e$
(``reactor neutrino anomaly''), and
iv) from the data of the radioactive source calibrations of the
GALLEX \cite{GALLEXcalib} and SAGE \cite{SAGEcalib}
solar neutrino experiments.
The evidences for sterile neutrinos from
the different data are typically at the level of up to
approximately $3\sigma$,
except in the case of the
LSND collaboration which claims a much higher C.L.

 Significant constraints on the parameters characterising the
oscillations involving sterile neutrinos follow from the negative
results of the searches for  $\nu_{\mu} \rightarrow \nu_e$ and/or
$\bar{\nu}_{\mu} \rightarrow \bar{\nu}_e$ oscillations
in the KARMEN  \cite{KARMEN2002}, NOMAD \cite{NOMAD2003},
ICARUS \cite{ICARUS13} and OPERA  \cite{OPERA13} experiments,
and from the nonobservation of effects of oscillations
into sterile neutrinos in the solar neutrino experiments and in the
studies of $\nu_{\mu}$ and/or $\bar{\nu}_{\mu}$ disappearance
in the CDHSW \cite{CDHSW84}, MINOS \cite{MINOSster}
and SuperKamiokande \cite{SuperKster} experiments.
Constraints on the number and masses of sterile neutrinos
are provided also by cosmological data
(see, e.g., \cite{Ade:2013lta,Mirizzi:2013kva,Wyman:2013lza}).
However, the constraints obtained so far cannot rule out
the possibility of existence of one or two light sterile
neutrinos which mix with the active flavour neutrinos.

 Two ``minimal'' phenomenological
models (or schemes) with light sterile neutrinos
are widely used in order
to explain the reactor neutrino and Gallium anomalies,
the LSND and MiniBooNE data as well as the results of
the negative searches for active-sterile neutrino
oscillations: the so-called ``$3 + 1$'' and ``$3 + 2$''
models, which contain respectively one and two sterile neutrinos
(right-handed sterile neutrino fields). The latter are assumed to
mix with the 3 active flavour neutrinos (left-handed flavour
neutrino fields)
(see, e.g., \cite{Kopp:2013vaa,Giunti:2013vaa}).
Thus, the ``$3 + 1$'' and ``$3 + 2$'' models have
altogether 4 and 5 light massive neutrinos
$\nu_j$, which in the minimal versions of these models are Majorana
particles. The additional neutrinos $\nu_4$ and $\nu_4$, $\nu_5$,
should have masses $m_4$ and $m_4$, $m_5$ at the eV scale (see
further). It follows from the data that if $\nu_4$ or $\nu_4$,
$\nu_5$ exist, they couple to the electron and muon in the weak
charged lepton current with couplings $U_{e k}$ and $U_{\mu k}$,
$k=4;~4,5$, which are approximately  $|U_{e k}|\sim 0.1$ and
$|U_{\mu k}|\sim 0.1$.

The hypothesis of existence of light sterile
neutrinos with eV scale masses and the indicated charged current
couplings to the electron and muon will be tested in a number of
experiments with reactor and accelerator neutrinos, and neutrinos
from artificial sources (see, e.g., \cite{SterNuWhitePaper,SnowM2013}
for a detailed list and
discussion of the planned experiments).

 In the present article we analyse the
possibility to test the hypothesis of
existence of light sterile neutrinos
which mix with the three active flavour neutrinos,
i.e., the existence of more than 3 light
massive Majorana neutrinos coupled to the electron
in the weak charged lepton current,
by studying the process of radiative emission of
neutrino pair (RENP) in atoms
\cite{RENP1} (see also, e.g., \cite{Fukumi:2012rn,Dinh:2012qb}
and references quoted therein).
The RENP is a process of  collective de-excitation of
atoms in a metastable level into emission
mode of a single photon plus a neutrino pair.
The process of RENP was shown to be sensitive
to the absolute values of the masses of the
emitted neutrinos, to the type of spectrum
the neutrino masses obey and
to the nature - Dirac or Majorana - of
massive neutrinos \cite{RENP1,Dinh:2012qb}.
If more than three light neutrinos
couple to the electron in the weak charged
lepton current and the additional neutrinos
beyond the three known have masses at the
eV scale, they will be emitted
in the RENP process. This will lead
to new observable features in the spectrum of the
photon, emitted together with the neutrino pair.
In the present article we analyse these features,
concentrating for simplicity on the $3 + 1$
phenomenological model with one sterile neutrino.

%
\section{One Sterile Neutrino: the $3 + 1$ Model}
\label{Sec:3plus1Model}
%

 We begin by recalling that
in the case of 3-neutrino mixing,
the sign of $\Delta m^2_{31(32)}$
cannot be determined from
the existing data and
the two possible signs of
$\Delta m^2_{31(2)}$,
as it is well known, correspond to two
types of neutrino mass spectrum.
In the widely used convention of numbering
the neutrinos $\nu_j$ with definite mass
in the two cases (see, e.g., \cite{PDG2012})
we shall also employ, the two spectra read:\\
{\it i) spectrum with normal ordering (NO)}:
$0 \leq m_1 < m_2 < m_3$, $\Delta m^2_{31(32)} >0$,
$\Delta m^2_{21} > 0$,
$m_{2(3)} = (m_1^2 + \Delta m^2_{21(31)})^{1\over{2}}$; \\~~
{\it ii) spectrum with inverted ordering (IO)}:
$0 \leq m_3 < m_1 < m_2$, $\Delta m^2_{32(31)}< 0$,
$\Delta m^2_{21} > 0$,
$m_{2} = (m_3^2 + \Delta m^2_{23})^{1\over{2}}$,
$m_{1} = (m_3^2 + \Delta m^2_{23} - \Delta m^2_{21})^{1\over{2}}$.\\
Depending on the value of the lightest neutrino mass,
${\rm min}(m_j)$, the neutrino mass spectrum can be:\\
%
{\it a) Normal Hierarchical (NH)}:
$m_1 \ll m_2 < m_3$, $m_2 \cong (\Delta m^2_{21})^{1\over{2}}
\cong 8.68 \times 10^{-3}$ eV,
$m_3 \cong (\Delta m^2_{31})^{1\over{2}}
\cong 4.97\times 10^{-2}~{\rm eV}$; or  \\
%
{\it b) Inverted Hierarchical (IH)}: $m_3 \ll m_1 < m_2$,
with $m_{1,2} \cong |\Delta m^2_{32}|^{1\over{2}}\cong 4.97\times 10^{-2}$ eV; or  \\
%
{\it c) Quasi-Degenerate (QD)}: $m_1 \cong m_2 \cong m_3 \cong m_0$,
$m_j^2 \gg |\Delta m^2_{31(32)}|$, $m_0 \gtap 0.10$ eV, $j=1,2,3$.

 We will be interested in the $3 + 1$ model, i.e., in the
possibility of existence of one extra sterile neutrino
beyond the three flavour neutrinos. In this case there will be
four massive Majorana neutrinos, $\nu_{1,2,3,4}$, with
$\nu_4$ being the heaviest neutrino,
$m_{1,2,3} < m_4$. Thus, the largest neutrino mass
squared difference in the case of the $3+1$ model with
NO (IO) 3-nutrino mass spectrum will be  $\Delta m^2_{41} >0$
($\Delta m^2_{43} >0$).

In the case of the $3+1$ scheme with NO neutrino mass spectrum,
$m_1 < m_2 < m_3 < m_4$, the masses $m_{2,3,4}$ can be
expressed in terms of the lightest neutrino mass $m_1$ and the three
neutrino mass squared differences $\Delta m^2_{21} > 0$, $\Delta
m^2_{31} > 0$ and $\Delta m^2_{41} >0$ as follows:
\be
\begin{split}
m_2  = \sqrt{m_{min}^2 + \Delta m^2_{21}}\,,~
m_3 = \sqrt{m_{min}^2 + \Delta m^2_{31}}\,,~
m_4 = \sqrt{m_{min}^2 + \Delta m^2_{41}}\,,~m_{min} \equiv m_{1}\,.
\end{split}
\ee
%

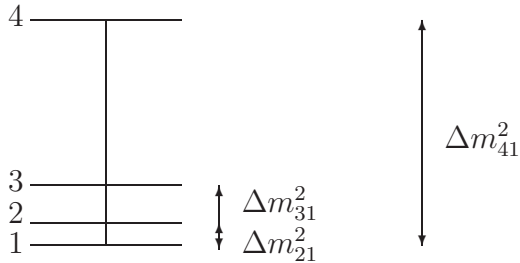
\begin{figure}
\unitlength=1mm
\begin{center}
\begin{picture}(100,40)
\put(20,5){\line(1,0){20}}
\put(20,8){\line(1,0){20}}
\put(20,13){\line(1,0){20}}
\put(20,35){\line(1,0){20}}
\put(30,5){\line(0,1){30}}
\put(45,4.5){\vector(0,1){3.5}}
\put(45,8){\vector(0,-1){3.5}}
\put(45,8){\vector(0,1){5}}
\put(72,5){\vector(0,1){30}}
\put(72,35){\vector(0,-1){30}}
\put(48,4){$\Delta m^2_{21}$}
\put(48,9.5){$\Delta m^2_{31}$}
\put(75,18){$\Delta m^2_{41}$}
\put(17,4){1}
\put(17,8){2}
\put(17,12.3){3}
\put(17,34.2){4}
\end{picture}
\end{center}
\caption{\label{specNH3plus1} The Mass spectrum in the $3 + 1$ NO (NH)
model.}
\end{figure}

\begin{figure}
\unitlength=1mm
\begin{center}
\begin{picture}(100,40)
\put(20,2){\line(1,0){20}}
\put(20,10){\line(1,0){20}}
\put(20,13){\line(1,0){20}}
\put(20,35){\line(1,0){20}}
\put(30,2){\line(0,1){33}}
\put(45,10){\vector(0,-1){8.5}}
\put(45,10){\vector(0,1){3.5}}
\put(45,13.5){\vector(0,-1){3.5}}
\put(72,5){\vector(0,1){30}}
\put(72,35){\vector(0,-1){34}}
\put(48,11){$\Delta m^2_{21}$}
\put(48,6){$\Delta m^2_{32}$}
\put(75,18){$\Delta m^2_{43}$}
\put(17,2){3}
\put(17,9){1}
\put(17,12.3){2}
\put(17,34.2){4}
\end{picture}
\end{center}
\caption{\label{specIH3plus1} The mass spectrum in the 3+1 IO (IH) model.}
\end{figure}
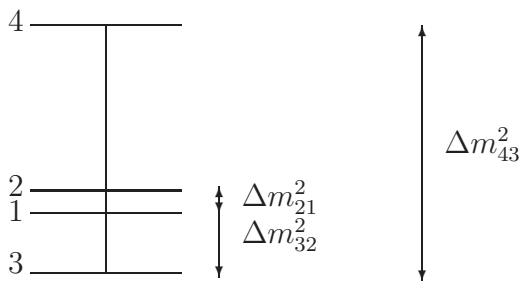

If the 3-neutrino mass spectrum of the $3 + 1$ scheme is of IO type,
the lightest neutrino mass is $m_{min} = m_3$,
i.e., we have $m_3 < m_1 < m_2 < m_4$,
$\Delta m^2_{21} > 0$, $\Delta m^2_{32} < 0$ and $\Delta m^2_{43} > 0$.
The masses $m_{1,2,4}$ are given by:
\be
\begin{split}
m_1 = & \sqrt{m_{min}^2 + |\Delta m^2_{32}| - \Delta  m^2_{21} }\,,~~
m_2 = \sqrt{m_{min}^2 + |\Delta m^2_{32}|}\,,~~
m_4 = \sqrt{m_{min}^2 + \Delta m^2_{43} }\,.
\end{split}
\ee
%
The mass spectra of the  $3 + 1$ NO (NH) and IO (IH) models are shown
schematically in Figs. \ref{specNH3plus1} and
\ref{specIH3plus1}, where the figures were taken
from \cite{Girardi:2013zra}.
%
%

 In the $3 + 1$ model there are four light massive
neutrinos and, correspondingly, the neutrino
mixing matrix - the
Pontecorvo, Maki, Nakagawa and Sakata (PMNS) matrix -
is a $4\times 4$ unitary matrix.
We will use the parametrisation of the PMNS matrix
adopted in \cite{Kopp:2013vaa}:
\be
U = O_{24} O_{23} O_{14} V_{13} V_{12}\,\,
diag (1,e^{i\alpha_{21}/2},e^{i\alpha_{31}/2},e^{i\alpha_{41}/2}),
\label{U3+1}
\ee
%
where $O_{ij}$ and $V_{kl}$ describe real and complex rotations in
$i-j$ and $k-l$ planes, respectively, and $\alpha_{21}$, $\alpha_{31}$ and
$\alpha_{41}$ are three CP violation (CPV) Majorana phases \cite{BHP80}.
Each of the $4\times 4$ matrices  $V_{12}$ and  $V_{13}$
contains one CPV phase, $\delta_{12}$ and $\delta_{13}$,
respectively, in their only two nonzero
nondiagonal elements:
%
\begin{eqnarray}
V_{ij}=D_{ij}\,O_{ij}(\theta_{ij})\,D_{ij}^*\,,
~~{(D_{ij})_{kr}}=0\,,~k\neq r\,,~~
{(D_{ij})_{kk}}=\left\{
\begin{array}{c}
e^{i\delta_{ij}}\,,~~~k=j; \\
~1~~~~~~k\neq j\,.  \\
\end{array}
\right.
\end{eqnarray}
%

 We did not include in $U$ a possible additional
matrix  $V_{34}$ corresponding to rotations in the
3-4 plane with angle $\theta_{34}$ (see, e.g., \cite{Kopp:2013vaa}),
i.e., we have set  $\theta_{34} = 0$, for simplicity.
The angle $\theta_{34}$, if nonzero, would be responsible for
direct $\nu_{\tau} \rightarrow \nu_s$ oscillations. At present
there do not exist data on this type of neutrino oscillations.
We will comment later on the effects of  $\theta_{34} \neq 0$
on the results of our analysis.

  In this study we will use two reference sets of  values of
the three sterile neutrino oscillation parameters
$\sin^2\theta_{14}$, $\sin^2\theta_{24}$
and $\Dmq_{41}$ ($\Dmq_{43}$)~
\footnote{The neutrino mass squared difference
$\Dmq_{41}$ ($\Dmq_{43}$) corresponds to
NO (IO) 3-neutrino mass spectrum.},
which are obtained in the analyses performed in
\cite{Kopp:2013vaa,Giunti:2013vaa}.
We will use the best fit values
\be
\sin^2\theta_{14}= 0.0225\,,~~\sin^2\theta_{24}= 0.0296\,,~~
\Delta m^2_{41(43)} = 0.93~{\rm eV^2}\,~~{\rm (A)}\,,
\label{A1}
\ee
%
found in \cite{Kopp:2013vaa} in the global analysis of all
the data (positive evidences and negative results)
relevant for the tests of the sterile neutrino hypothesis.
 In ref. \cite{Kopp:2013vaa} a combined constraint on
$\sin^2\theta_{24}$ and
$\sin^2\theta_{34}$ was also obtained from the global analysis of
the $\nu_{\mu}$ disappearance data (see Fig. 5 (middle panel) in
\cite{Kopp:2013vaa}). For the best fit value
of $\sin^2\theta_{24}= 0.0296$ and $\Delta m^2_{41(43)} = 1.0~{\rm eV^2}$
this constraint implies $\sin^2\theta_{34} \ltap 0.05$ at 99\% C.L.
In what follows we will present results both for
$\sin^2\theta_{34} = 0$ and for $\sin^2\theta_{34} = 0.05$.

 Global analysis of the sterile neutrino related data
was performed, as we have already noticed,
also in \cite{Giunti:2013vaa} (for earlier
analyses see, e.g., \cite{Archi12}). The authors of
\cite{Giunti:2013vaa} did not include in the data set used the
MiniBooNE results at $E_{\nu} \leq 0.475$ GeV, which show an excess
of events over the estimated background \cite{MBooNEexcess}. The
nature of this excess is not well understood at present. For the
best values of $\sin^2\theta_{14}$ and
$\Delta m^2_{41(43)}$ the authors of \cite{Giunti:2013vaa} find:
 \be
\sin^2\theta_{14}= 0.0283\,,~~\sin^2\theta_{24}= 0.0124\,,~~
\Delta m^2_{41(43)} = 1.60~{\rm eV^2}\,~~{\rm (B)}\,.
\label{B1}
\ee
%
The quoted  values of $\sin^2\theta_{14}$  and
$\Delta m^2_{41(43)}$
are close to the best fit values found in
\cite{Kopp:2013vaa} in the analysis of the
$\nu_e$ and $\bar{\nu}_e$ disappearance data:
$\sin^2\theta_{14}= 0.023$, $\Delta m^2_{41(43)} = 1.78~{\rm eV^2}$.
The authors of ref. \cite{Giunti:2013vaa} give also
the allowed ranges of values of $\Delta m^2_{41(43)}$ and
$\sin^2\theta_{14,24}$ at various confidence levels.
However, taking into account the uncertainties in the
values of $\Delta m^2_{41(43)}$, $\sin^2\theta_{14}$ and
$\sin^2\theta_{24}$ is beyond the scope
of the present study.

  In what concerns the 3-neutrino oscillations parameters
$\sin^2\theta_{12}$, $\sin^2\theta_{23}$
$\sin^2\theta_{13}$ and $\Delta m^2_{31(32)}$,
in our numerical analysis we will use the their best
fit values found in \cite{Capozzi:2013csa}:
\begin{eqnarray}
\label{dmq2131}
&\Delta m^2_{21} = 7.54 \times 10^{-5} \ {\rm eV^2}\,,~~
|\Delta m^2_{31(32)}| = 2.47~(2.42) \times 10^{-3} \ {\rm eV^2}\,,\\
\label{th122313}
&\sin^2\theta_{12} = 0.308\,,~
\sin^2\theta_{13} = 0.0234~(0.0240)\,,~\sin^2\theta_{23} = 0.437~(0.455)\,,
\end{eqnarray}
%
where the values (the values in brackets)
correspond to NO (IO) neutrino mass spectrum.

 It should be added that global analyses of the
neutrino oscillation data relevant for the test
of the sterile neutrino hypothesis (positive evidences and
the negative results) in the
3 + 1 scheme of interest, in which
the 3-neutrino mixing parameters
are treated as free parameters as well,
have not been performed so far.
Since the sterile neutrino mixing angles are
of the order of 0.1 and they will affect the 3-neutrino mixing
angles, in particular, via the unitarity conditions,
$\sum_j  |U_{lj}|^2 = 1$, $l=e,\mu$,
one can expect naively that the changes, e.g., in
$|U_{lj}|^2$, $j=1,2,3$, where $U_{lj}$ are the elements
of the first and second rows of the neutrino mixing  matrix,
to be of the order of 0.01. Since
$\sin^2\theta_{13} = 0.0234$, one might expect, in particular,
sizable effects on the value of $\sin^2\theta_{13}$.
However, the detailed study performed in
ref.  \cite{Kopp:2013vaa} showed that actually the
value of $\sin^2\theta_{13}$ as determined in
3 active neutrino oscillation analysis remains
stable with respect to the presence of sterile neutrinos.
The values of the other neutrino mixing parameters, relevant
in our analysis, $\sin^2\theta_{12} \cong 0.31 >> 0.01$
and $\sin^2\theta_{23} \cong 0.44~(0.45) >> 0.01$,
and the effect of the presence of sterile neutrinos
in the mixing is negligible for them.

 Finally, in the analysis we will perform, the two Dirac
and three Majorana CPV phases will be varied in their entire
defining intervals.

%
\section{The Process of RENP Involving
Sterile Neutrinos}
\begin{figure*}[htbp]
 \begin{center}
 \epsfxsize=0.4\textwidth
 \centerline{\epsfbox{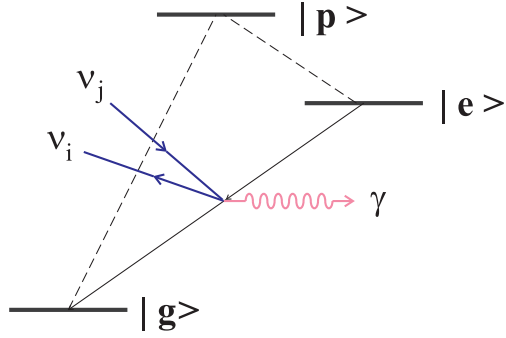}}
  \hspace*{\fill}
   \caption{
Atomic level scheme (of $\Lambda-$type \cite{Dinh:2012qb})
for RENP
$|e \rangle \rightarrow |g\rangle + \gamma + \nu_i\nu_j$,
$\nu_i$ and $\nu_j$ being neutrino mass eigenstates.
}
\label{RENP}
 \end{center}
\end{figure*}

%
%
  For a single atom, the process of
radiative emission of neutrino pair (RNEP) of interest
is $|e\rangle \rightarrow |g\rangle + \gamma +(\nu_i + \nu_j)$,
where $\nu_i$'s are the neutrino mass eigenstates
(see Fig. \ref{RENP}). In the case
%
\noindent of interest we have $i,j=1,2,3,4$.
If $\nu_i$ are Dirac fermions, $(\nu_i + \nu_j)$
should be understood as a pair of
neutrino anti-neutrino with masses $m_i$ and $m_j$,
respectively. If neutrinos with definite mass
are Majorana particles,
we have $\bar{\nu}_i \equiv \nu_i$ and $(\nu_i + \nu_j)$
are the Majorana neutrinos with masses $m_i$ and $m_j$.
The proposed experimental method is to measure,
under irradiation of two counter-propagating
trigger lasers, the continuous photon ($\gamma$)
energy spectrum below each of the thresholds
$\omega_{ij}$ corresponding to the production of different
pairs of neutrinos, $\nu_1\nu_1$, $\nu_1\nu_2$, $\nu_2\nu_2$,...,
$\omega < \omega_{ij}$,
$\omega$ being  the photon energy, and \cite{RENP1,RENP2}
\begin{eqnarray}
&&
\omega_{ij} = \omega_{ji}
= \frac{\epsilon_{eg}}{2} - \frac{(m_i+m_j)^2}{2\epsilon_{eg}}\,,~~~
i,j=1,2,...,~~m_{i},m_j\geq 0\,,
\label{thresholds}
\end{eqnarray}
%
where $\epsilon_{eg}$ is the energy difference between
the two relevant atomic levels.
For four massive neutrinos there are altogether
10 different pairs $\nu_1\nu_1$, $\nu_1\nu_2$,...,
$\nu_3\nu_4$, $\nu_4\nu_4$, and, correspondingly,
10 threshold energies $\omega_{ij}$.
The disadvantage of the method is the smallness of
the RENP rate, which
is proportional to $G_F^2$, $G_F = 10^{-23} {\rm eV^{-2}}$.
This can possibly be overcomed by  ``macro-coherence''
amplification of the rate
\cite{macro-coherence,psrdynamics},
the amplification factor being
$\propto n^2V$, where n is the number density
of excited atoms and V is the
volume irradiated by the trigger laser.
For $n$ at the order of  $N_A/cm^3$, where $N_A$ is Avogadro's number,
and $V\sim 100~{\rm cm^3}$, the rate is observable.
The macro-coherence of interest is developed by
irradiation of two trigger lasers of
frequencies $\omega_1$, $\omega_2$,
satisfying $\omega_1+\omega_2=\epsilon_{eg}$. It is a
complicated dynamical process.
The asymptotic state of fields and target atoms in the latest
stage of trigger irradiation is described by a static solution
of the master evolution equation.
In many cases there is a remnant state consisting of field
condensates (of the soliton type) accompanied
with a large coherent medium polarisation.
This asymptotic target state is stable against two photon
emission (except for minor ``leakage'' from the edges of the
target), while RENP occurs from any point in the target
\cite{macro-coherence,psrdynamics}. A Group at Okayama University,
Okayama, Japan, is working on the experimental realisation of
the macro-coherent RENP \cite{Fukumi:2012rn}.

   As was indicated above, the physical observable
of interest in the process of RENP, that, in principle,
can be measured experimentally,
is the single photon spectrum.
The features of the phton spectrum in the
case of 3-neutrino mixing, which allow to get information
about the absolute neutrino mass scale, the neutrino mass
spectrum and about the nature - Dirac or Majorana -
of massive neutrinos, were discussed in detail on the
example of a specific (combined $E1\times M1$) atomic
transition in \cite{Dinh:2012qb}. Here we will generalise
the results obtained in \cite{Dinh:2012qb}
to the case of the $3 + 1$ scheme with one sterile
neutrino and four massive Majorana neutrinos.
We will be primarily interested in the new
photon spectrum features associated with the 4th massive
Majorana neutrino related to the presence of the sterile
neutrino in the model. A detailed discussion of the atomic
physics aspects of the problem is given, e.g.,
in \cite{Fukumi:2012rn}.

 The photon spectrum of interest,
or more precisely, the photon spectral
rate, i.e., the rate of number of events per unit
time at each photon energy $\omega$,
in the cases of  transitions in atoms considered
in refs. \cite{Dinh:2012qb} (see further),
can be written as:
\begin{eqnarray}
&&
\Gamma_{\gamma 2\nu}(\omega)  = \Gamma_0 I(\omega) \eta(t)
\,, \hspace{0.5cm}
\label{spectrum}
\\
&&
I(\omega) = \frac{1}{(\epsilon_{pg}-\omega)^2}
\sum_{ij}\,|a_{ij}|^2\,\Delta_{ij}(\omega)
\left ( I_{ij}(\omega) -
\delta_M m_i\, m_j\,B^M_{ij}\,
\right )
\,,
\label{spectrum00}
\\ &&
B^M_{ij} = 
\left (1 - 2\,\frac{({\rm Im}(a_{ij}))^2}{|a_{ij}|^2}\right )
\,,~~~
\label{spectrum01}
\\ &&
\Delta_{ij}(\omega)
=\frac{1}{\epsilon_{eg} (\epsilon_{eg} -2\omega) }\left\{
\left( \epsilon_{eg} (\epsilon_{eg} -2\omega) - (m_i + m_j)^2\right)
\left( \epsilon_{eg} (\epsilon_{eg} -2\omega) - (m_i - m_j)^2\right)
\right\}^{1/2}
\,,
\label{spectrum3}
\\ &&
I_{ij}(\omega) =\left(
\frac{1}{3}\epsilon_{eg}(\epsilon_{eg}-2\omega)
+ \frac{1}{6}\omega^2
- \frac{1}{18}\omega^2 \Delta_{ij}^2(\omega)
- \frac{1}{6}(m_i^2+ m_j^2)
- \frac{1}{6}\frac{(\epsilon_{eg}-\omega)^2}
{\epsilon_{eg}^2(\epsilon_{eg}-2\omega)^2}(m_i^2 - m_j^2)^2
\right)
\,,
\nonumber  \\ &&
\label{spectrum1}
\end{eqnarray}
%
Here $\Gamma_0$ determines the overall rate of the process and
$\eta_{\omega}(t)$ is a dynamical dimensionless factor.
Explicit expressions for both $\Gamma_0$ and $\eta_{\omega}(t)$
are given in \cite{Dinh:2012qb}. Following the discussion
in \cite{Dinh:2012qb}
(see also \cite{RENP2}),
the factor $\eta_{\omega}(t)$ will be set to unity.
Further, the factor $|a_{ij}|^2$ in the spectral function
$I(\omega)$ is given in the case of
the ``3+1'' scheme of interest by:
\begin{equation}
|a_{ij}|^2 =
\left | U_{ei}^*\,U_{ej} -
\frac{1}{2}\sum_{l=e,\mu,\tau}  U_{l i}^*\,U_{l j} \right|^2 =
\left | U_{ei}^*\,U_{ej} -
\frac{1}{2}\left (\delta_{ij} -  U_{s i}^*\,U_{s j}\right )\right|^2\,,~
i,j=1,...,4\,,
\label{aij}
\end{equation}
%
where  $U_{s i}$ and $U_{s j}$ are two elements of the 4th row
of the PMNS matrix.
In the limit of  $U_{s i} = 0$, $i=1,2,3,4$, and for $i,j$ taking values $i,j=1,2,3$,
the expression for $|a_{ij}|^2$, eq. (\ref{aij}),
and for $I(\omega)$, eqs. (\ref{spectrum00}) - (\ref{spectrum1}),
reduce to those corresponding to the reference scheme of mixing of
3 active flavour neutrinos with three light massive neutrinos.
The term $\propto \delta_M m_i m_jB^M_{ij}$ in $I(\omega)$
appears only in the Majorana neutrino case:
$\delta_M = 1$ ($\delta_M = 0$) if $\nu_j$ are
Majorana (Dirac) particles.
Let us add that, more generally, the term
$\propto m_im_j(1 - 2({\rm Im}(a_{ij}))^2/|a_{ij}|^2)$
is similar to, and has the same physical origin as,
the term $\propto M_iM_j$ in the production cross section
of two different Majorana neutralinos
$\chi_i$ and $\chi_j$
with masses $M_i$ and $M_j$ in the
process of $e^- + e^+ \rightarrow \chi_i + \chi_j$
\cite{eechi1chi286}. The term $\propto M_iM_j$
of interest determines, in particular, the threshold behavior
of the indicated cross section.

 In the limit of massless
neutrinos the spectral rate becomes
\begin{eqnarray}
&&
I(\omega; m_i = 0) =
\frac{\omega^2 - 6\epsilon_{eg}\omega +
3\epsilon_{eg}^2}{12(\epsilon_{pg}-\omega)^2}
\,,
\label{Iomegamassless}
\end{eqnarray}
%
where the prefactor of  $\sum_{ij}|a_{ij}|^2 =3/4$  is
calculated using the unitarity of the neutrino mixing matrix.

In what follows we will perform a numerical
analysis for the case of Yb atom
and energy levels relevant for
the RENP process of interest,
for which a similar analysis was
performed in the case of mixing of
three active flavour neutrinos
in \cite{Dinh:2012qb}:
\begin{eqnarray}
&&
{\rm Yb}; \hspace{0.3cm}
|e \rangle = (6s6p)\,^3P_0
\,, \hspace{0.3cm}
|g \rangle = (6s^2)\, ^1S_0
\,, \hspace{0.3cm}
|p \rangle = (6s6p)\,^3P_1\,.
\end{eqnarray}
%
The atomic energy differences are \cite{nist}:
\begin{eqnarray}
&&
\epsilon_{eg} =  2.14349~{\rm eV}
\,, \hspace{0.3cm}
\epsilon_{pg} = 2.23072~{\rm eV}
\,.
\label{AtomicEnergy}
\end{eqnarray}
%
In the case of Yb atom considered,
the overall rate factor $\Gamma_0$ is given by
\begin{eqnarray}
&&
\Gamma_0 \sim 0.37~{\rm mHz} (\frac{n}{10^{21}~{\rm cm}^{-3}})^3
\frac{V}{10^2~{\rm cm}^3}\,,
\end{eqnarray}
%
where the number is valid for the Yb first excited state of $J=0$
\footnote{
If one chooses the other intermediate path, $^1P_1$,
the rate $\Gamma_0$ is estimated to
be of order, $1\times 10^{-3}$ mHz,
a value much smaller than that of the $^3P_1$ path.}.

  In the present article we concentrate on the elementary particle physics
potential of the proposed method to get information about the
neutrino masses, neutrino mixing and the nature of massive neutrinos,
and, more specifically, about the existence of more than three
light massive neutrinos related to the exisatence of sterile neutrinos.
The technical aspects of the possible experiment based on the method
considered are discussed in, e.g., \cite{Fukumi:2012rn}.

%
\section{The Photon Spectrum Features in the Case of
the ``3 + 1'' Scheme}
%
%

  It follows from eqs. (\ref{spectrum}) and (\ref{spectrum00})
that the rate of emission of a given pair of neutrinos
$(\nu_i + \nu_j)$ is suppressed, in particular,
by the factor $|a_{ij}|^2$, independently of the nature
of massive neutrinos. In the case of mixing of 3 active
flavour neutrinos and 3 massive neutrinos $\nu_j$, $j=1,2,3$,
the 6 factors, corresponding to
the emission of the 6 different neutrino pairs,
$|a_{ij}|^2$, $i,j=1,2,3$, do not depend
on the CPV phases in the $3\times 3$
unitary PMNS matrix; they depend only
on the values of the angles $\theta_{12}$ and $\theta_{13}$.
The expressions for these 6 different factors
$|a_{ij}|^2$, $i,j=1,2,3$,
in terms of the sines and cosines of
the mixing angles $\theta_{12}$ and $\theta_{13}$,
are given in \cite{Dinh:2012qb}, where
the values of $|a_{ij}|^2$ corresponding to the
best fit values of $\sin^2\theta_{12} =0.307$ and
$\sin^2\theta_{13} = 0.0241~(0.0244)$,
obtained in the global analysis in \cite{Fogli:2012XY},
are also quoted
(see Table 1 in \cite{Dinh:2012qb}).
\begin{table}
\centering
\caption{
\label{tab_aijA}
The quantity $|a_{ij}|=|U^*_{ei}U_{ej}-\frac{1}{2}\sum_{l=e,\mu,\tau}  U_{l i}^*U_{l j}|$ (NO), case A}
    \begin{tabular}{|c|c|c|c|c|}
        \hline
        $|a_{11}|$      & $|a_{12}|$      & $|a_{13}|$      & $|a_{14}|$      & $|a_{22}|$      \\ \hline
        $0.1612-0.1822$ & $0.4375-0.4471$      & $0.1136-0.1350$      & $0.0208-0.1047$ & $0.1864-0.2058$ \\ \hline
        $|a_{23}|$      & $|a_{24}|$      & $|a_{33}|$      & $|a_{34}|$      & $|a_{44}|$      \\  \hline
        $0.0731-0.0941$      & $0.0057-0.0989$ & $0.4680-0.4731$ & $0.0431-0.0664$ & $0.0032$ \\ \hline
    \end{tabular}
   \end{table}
\begin{table}
\centering \caption{\label{tab_aijB}
The quantity $|a_{ij}|=|U^*_{ei}U_{ej}-
\frac{1}{2}\sum_{l=e,\mu,\tau}  U_{l i}^*U_{l j}|$ (NO), case B}
    \begin{tabular}{|c|c|c|c|c|}
        \hline
        $|a_{11}|$      & $|a_{12}|$      & $|a_{13}|$      & $|a_{14}|$      & $|a_{22}|$      \\ \hline
        $0.1600-0.1753$ & $0.4395-0.4465$      & $0.1176-0.1308$      & $0.0417-0.0963$ & $0.1937-0.2076$ \\ \hline
        $|a_{23}|$      & $|a_{24}|$      & $|a_{33}|$      & $|a_{34}|$      & $|a_{44}|$      \\  \hline
        $0.0776-0.0891$      & $0.0089-0.0831$ & $0.4724-0.4761$ & $0.0228-0.0485$ & $0.0081$ \\ \hline
    \end{tabular}
   \end{table}
%

 In the case of the ``3 + 1'' scheme of interest with
4 massive neutrinos, there are altogether
10 factors $|a_{ij}|^2$, $i,j=1,2,3,4$, corresponding to
the emission of the 10 different neutrino pairs.
Moreover, in this case the factors of interest
depend, in particular,
on the  Dirac and Majorana CPV phases present in the
$4\times 4$ unitary PMNS matrix
\footnote{It follows from eqs. (\ref{U3+1})
and (\ref{aij}), in particular, that
the qunatities $|a_{ii}|$, $i=1,2,3$,
do not depend on the Majorana phases;
they depend on the Dirac phases
$\delta_{12}$ and $\delta_{13}$
only through the term  $|U_{s i}|^2$.
}
given in eq. (\ref{U3+1}).
By using the best fit values of the sterile neutrino
mixing parameters in the cases A and B,
quoted in eqs. (\ref{A1}) and (\ref{B1}),
the 3-neutrino mixing parameters for NO neutrino mass
spectrum given in eq. (\ref{th122313}),
and varying the Dirac and Majorana phases in the
interval  $[0,2\pi]$, we have obtained
the intervals of values in which
the factors $|a_{km}|$, $k,m=1,2,3,4$,
lie in the case of NO spectrum.
The results are given in
Tables \ref{tab_aijA} and \ref{tab_aijB}.
Performing a similar calculation
assuming IO neutrino mass spectrum
we found that the results for
$|a_{km}|$, $k,m=1,2,3,4$,
differ from those obtained
for the NO spectrum by $\sim 10^{-3}$.
Comparing the values of the factors
$|a_{i,j}|$, $i,j=1,2,3$, quoted in
Table 1 in \cite{Dinh:2012qb},
with those given in Tables \ref{tab_aijA} and \ref{tab_aijB}
we can conclude that the factors
$|a_{i,j}|$ in the ``3 + 1 '' scheme,
which do not involve the 4th neutrino,
i) change relatively little
when one varies the two Dirac and
the three Majorana CPV phases,
and ii) can differ at most by approximately 0.02
from the factors in the 3-flavour neutrino mixing scheme.
The same conclusion is valid in the case
of IO neutrino mass spectrum.
This implies that the presence of a 4th (sterile) neutrino
in the mixing in the ``3 + 1'' scheme
has a little effect on the sensitivities of the RENP process
to the masses and the mixing of the three lighter neutrinos
$\nu_{1,2,3}$, i.e., the absolute neutrino
mass scale, the type of the neutrino
mass spectrum and the nature of massive neutrinos,
associated with the sub-mixing of
the 3 active neutrinos, which were analised in detail
in \cite{Dinh:2012qb}.
 This conclusion remains valid also
for $\sin^2\theta_{34} = 0.05$ in the case A
of values of the sterile neutrino
oscillation parameters.
The factors $|a_{i,j}|$, $i,j=1,2,3,4$,
in this case are given in Table \ref{tab_aijA2}.
Comparing the values of $|a_{i,j}|$
in Tables \ref{tab_aijA} and \ref{tab_aijA2}
we see that  a non-zero $\sin^2\theta_{34} = 0.05$
leads to:
i) a change of  $|a_{i,j}|$,
$i,j=1,2,3$, by at most 0.01, except
for the case $i=j=3$, in which the change
is approximately by 0.03;
ii) a change in the minimal and maximal values
of  $|a_{14}|$ by 0.02, of $|a_{24}|$ - approximately
by 0.02 and 0.035,  and of the value of $|a_{44}|$ -
approximately by 0.02. The largest is the change
of the interval of values of $|a_{34}|$ -
it is shifted to larger values by approximately
0.07.

%
\begin{table}
\centering \caption{
\label{tab_aijA2}
The quantity $|a_{ij}|=|U^*_{ei}U_{ej}-\frac{1}{2}\sum_{l=e,\mu,\tau}  U_{l i}^*U_{l j}|$ (NO), case A, $\sin^2{\theta_{34}}=0.05$}
    \begin{tabular}{|c|c|c|c|c|}
        \hline
        $|a_{11}|$      & $|a_{12}|$      & $|a_{13}|$      & $|a_{14}|$      & $|a_{22}|$      \\ \hline
        $0.1631-0.1744$ & $0.4425-0.4493$      & $0.1019-0.1371$      & $0.0431-0.0883$ & $0.1987-0.2053$ \\ \hline
        $|a_{23}|$      & $|a_{24}|$      & $|a_{33}|$      & $|a_{34}|$      & $|a_{44}|$      \\  \hline
        $0.0667-0.0930$      & $0.0239-0.0638$ & $0.4329-0.4450$ & ${0.1185-0.1429}$ & $0.0269$ \\ \hline
    \end{tabular}
   \end{table}

Below we give a brief summary of the
RENP phenomenology of the
3-flavour neutrino sub-mixing scheme,
which practically coincides with the phenomenology
and the results obtained in \cite{Dinh:2012qb} by
analysing the reference 3-flavour neutrino mixing
scheme with three massive neutrinos.
%
%
\subsection{The  3-Flavour Neutrino Sub-Mixing:
A Brief Summary of the RENP Phenomenology
}
%
%

\begin{figure}
 \begin{center}
\vskip -0.4cm
\includegraphics[width=9.5cm,height=6.5cm]{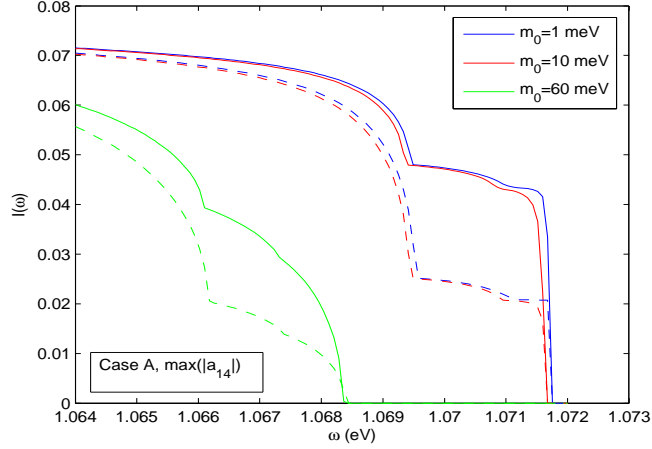}
   \caption{Photon energy spectra from the
considered $^3P_0 \rightarrow {^1S_0}$ transition in Yb
in the region of the 3-neutrino sub-mixing
thresholds $\omega_{ij}$, $i,j=1,2,3$,
in the  ``3 + 1'' scheme with 4 massive Dirac neutrinos,
for three sets of values of neutrino masses
corresponding to $m_0 = 1, 10, 60 {\rm meV}$ and for
NO (solid lines) and IO (dashed lines) neutrino mass spectrum.
The spectra are obtained for the set A of values of neutrino
oscillation parameters and for values of the CPV phases
which maximise the factor $|a_{14}|$.
}
\label{Yb3nuDiracNOIO}
 \end{center}
\end{figure}
\begin{figure}
 \begin{center}
\begin{tabular}{cc}
\includegraphics[width=9.5cm,height=6.5cm]{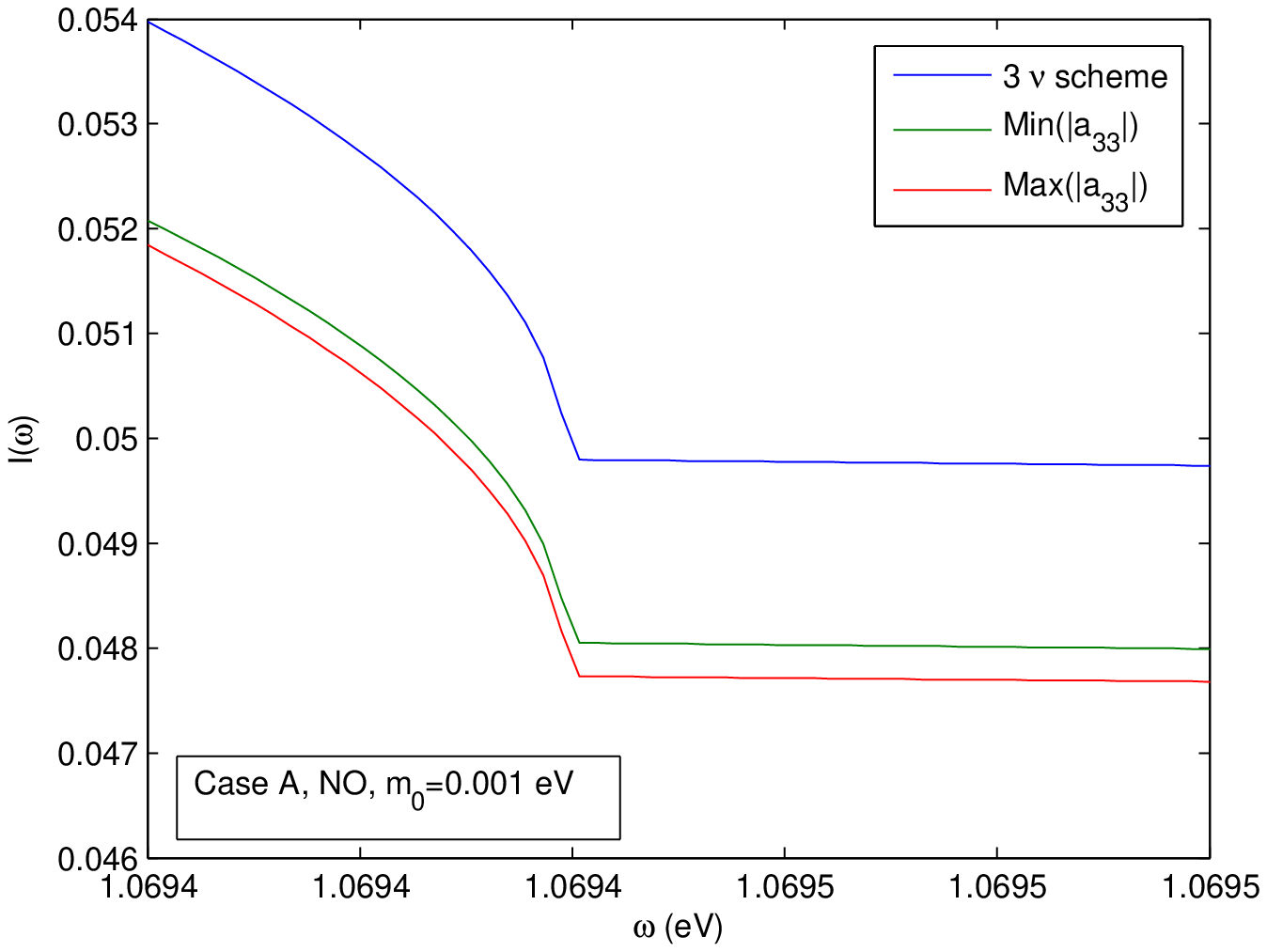} &
\includegraphics[width=7.5cm,height=6.5cm]{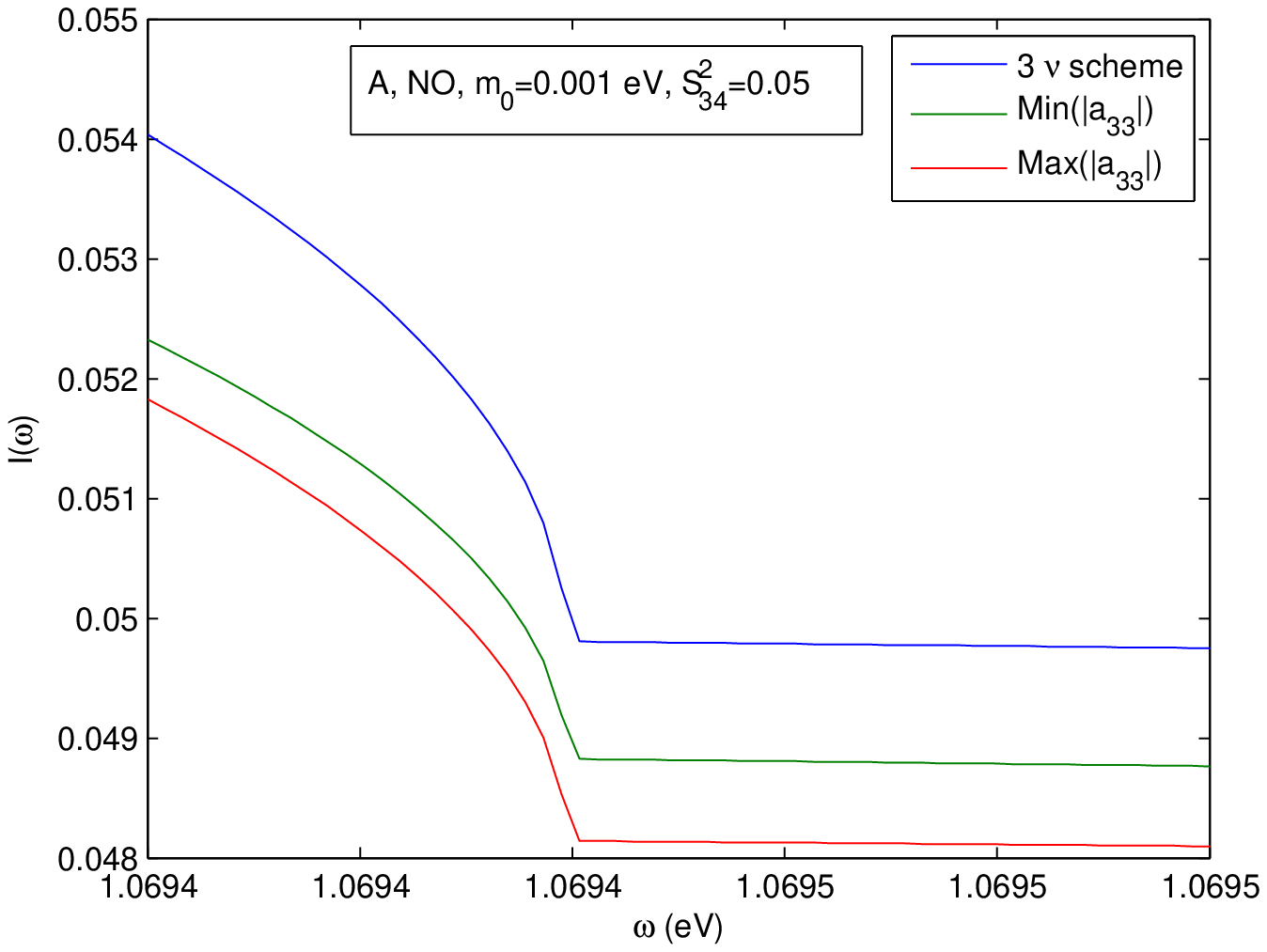}
\end{tabular}
 \caption{Photon energy spectrum from the
$^3P_0 \rightarrow {^1S_0}$ transition in Yb
in the region of the
 $(\nu_3 +\nu_3)$ emission
threshold $\omega_{33}$ in the cases of 3-neutrino
and ``3+1''-neutrino  mixing with
massive Dirac neutrinos, for $m_0 = 1~{\rm meV}$,
and NO neutrino mass spectrum. In the case of
``3+1''-neutrino  mixing results in the case A for
$\sin^2\theta_{34} = 0$ (left panel) and 0.05
(right panel) and for
the  maximal and minimal values of $|a_{33}|$
are shown.
}
\label{Yb33NO34}
 \end{center}
\end{figure}

\begin{figure}
 \begin{center}
\begin{tabular}{cc}
\includegraphics[width=9.5cm,height=6.5cm]{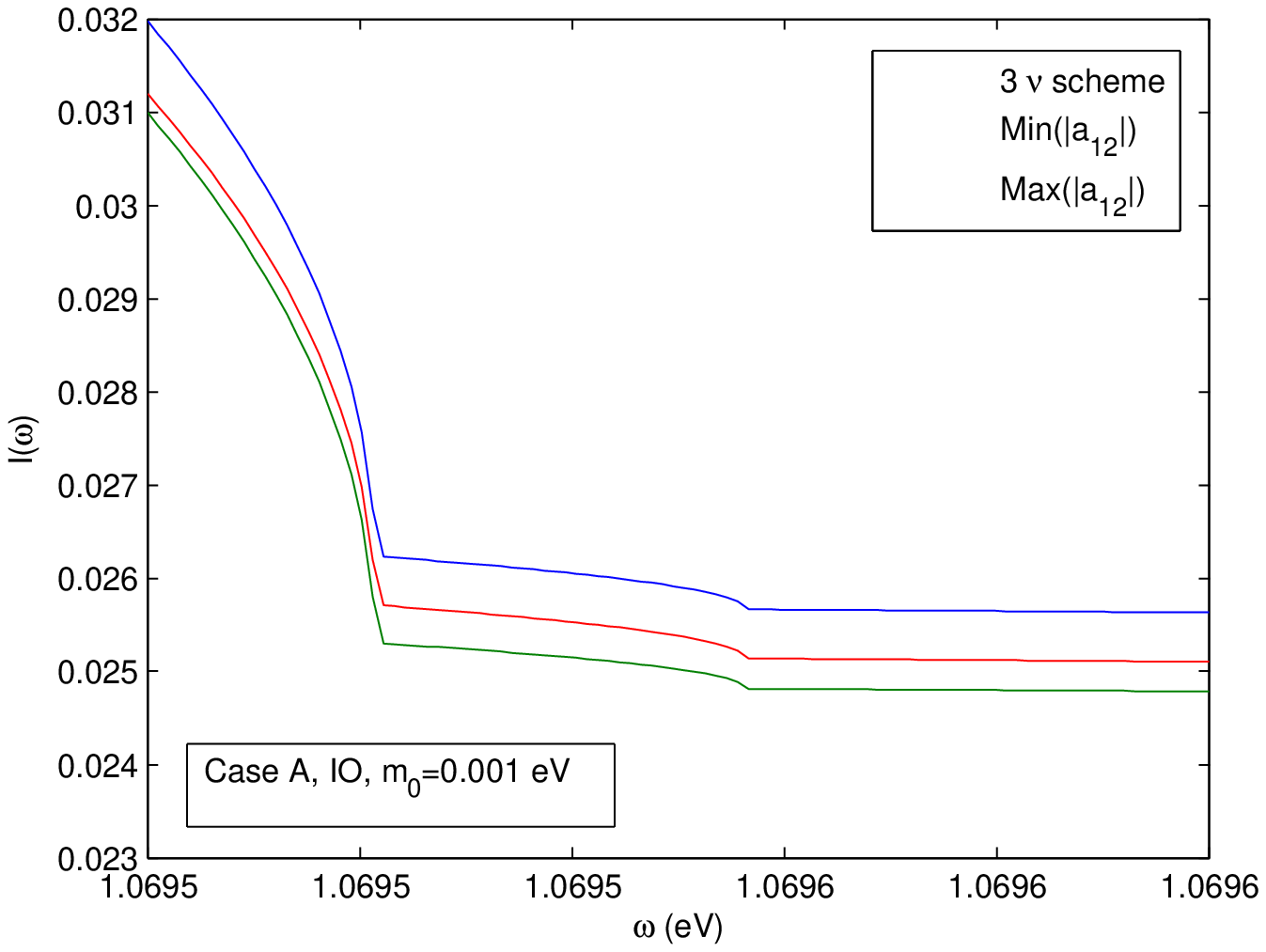} &
\includegraphics[width=7.5cm,height=6.5cm]{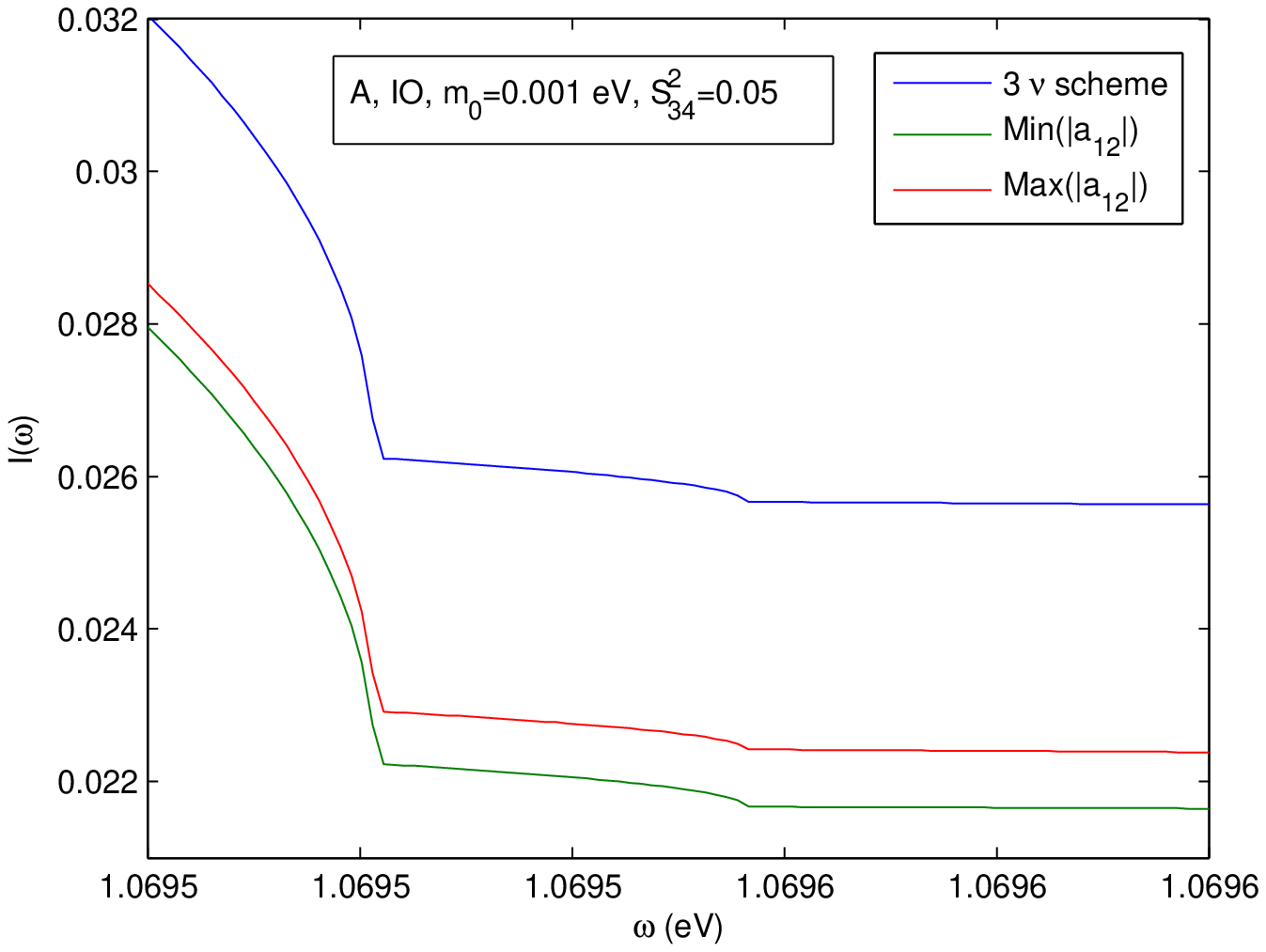}
\end{tabular}
\caption{The same as in Fig. \ref{Yb33NO34} but
for IO neutrino mass spectrum and the
$(\nu_1 +\nu_2)$ emission threshold. In the case of
``3+1''-neutrino  mixing results for
the  maximal and minimal values of $|a_{12}|$
are shown.
}
\label{Yb12IO34}
 \end{center}
\end{figure}
%

All the neutrino physics information
on the 3 light neutrino sub-mixing
of interest is contained in the dimensionless spectral
function $I(\omega)$ for values of $\omega$
near the thresholds $\omega_{ij}$, $i,j,=1,2,3$.
This is illustrated in Fig. \ref{Yb3nuDiracNOIO}
in which we show  $I(\omega)$ for values of
$\omega \leq \omega_{ij}$, $i,j,=1,2,3$,
in the case of three massive Dirac neutrinos
for three different sets of values of the neutrino
masses  (corresponding to the smallest mass $m_{0}=1,\; 10,\; 60$ meV)
and for both the NO ($\Delta m^2_{31(32)} > 0$) and
IO ($\Delta m^2_{31(32)} < 0$) neutrino mass spectra.
We note that the locations of the thresholds
corresponding to the three values of $m_0$
(and that can be seen in the figure)
differ substantially.
This feature can be used to
determine the absolute neutrino mass scale, i.e.,
the smallest neutrino mass, as evident in
differences of spectrum shapes for different masses of
$m_0$, 1,~10,~60 meV in Fig. \ref{Yb3nuDiracNOIO}.
In particular, the smallest mass can be determined by
locating the highest threshold ($\omega_{11} $ for NO and
$\omega_{33} $ for IO).
Also the location of the most prominent kink, which
is due to the heavier neutrino pair emission thresholds
($\omega_{33}$ in the NO case
and $\omega_{12}$
in the IO case), can independently be used to
extract the smallest neutrino mass value.
If the spectrum is of the NO type, the measurement
of the position of the kink will determine the
value of $\omega_{33}$ and therefore of $m_3$.
For the IO spectrum, the threshold  $\omega_{12}$ is very
close to the thresholds $\omega_{22}$  and $\omega_{11}$.
The rates of emission of the pairs $(\nu_2 + \nu_2)$
and $(\nu_1 + \nu_1)$, however, are smaller approximately
by the factors 10.0 and 13, respectively,
than the rate of emission of $(\nu_1 + \nu_2)$.
Thus, the kink due to the
$(\nu_1 + \nu_2)$ emission will be the easiest to observe.
The position of the kink will allow to determine
$(m_1 + m_2)^2$ and thus the absolute neutrino
mass scale.

  Once the absolute neutrino mass scale is determined,
the distinction between the NO (NH) and IO (IH)
spectra can be made by measuring the ratio of
rates below and above the thresholds
$\omega_{33}$ and $\omega_{12}$ (or $\omega_{11}$),
respectively. For $m_0 \ltap 20$ meV and NH (IH) spectrum, for instance,
the ratio of the rates at $\omega$ just above the
$\omega_{33}$ ($\omega_{12}$) threshold and sufficiently
far below the indicated thresholds, $\tilde{R}$,
is $\tilde{R} \cong 0.70$
in the case of NH spectrum, and $\tilde{R} \cong 0.36$
if the spectrum is of the IH type (see also  \cite{Dinh:2012qb}).
As Fig.\ref{Yb3nuDiracNOIO} indicates,
this ratio changes little
when $m_0$ increases up to $m_0 \cong 100$ meV.

  The effect of the presence of
the 4th neutrino $\nu_4$ in the mixing
on the photon spectrum in the region of the
$\nu_3 +\nu_3$ ($\nu_1 +\nu_2$) emission
threshold $\omega_{33}$  ($\omega_{12}$)
in the case of the ``3+1'' scheme,
NO (IO) neutrino mass spectrum with
$m_0 = 1$ meV, $\sin^2\theta_{34} =0;~0.05$
 and Dirac neutrinos $\nu_j$, is illustrated
in Fig. \ref{Yb33NO34} (Fig. \ref{Yb12IO34}).
As Fig. \ref{Yb33NO34} (Fig. \ref{Yb12IO34}) indicates,
the presence of the 4th neutrino leads to an overall
decreasing of the photon  spectrum
near the  $\omega_{33}$  ($\omega_{12}$) threshold
by approximately 0.002 (0.001 - 0.003)
in the NO (IO) case with respect to the spectrum
corresponding to the 3-neutrino mixing.
The quoted magnitude of the change of the spectrum
remains practically the same when $m_0$
is increased up to 0.10 eV. In the case of IO spectrum it is maximal
for $\sin^2\theta_{34} = 0.05$.
The observation of the indicated relatively
small difference
between the two photon spectra under discussion,
corresponding  the 3-neutrino and
the (3+1)-neutrino mixing,
requires a high precision measurement
of the photon spectrum.

 Determining the nature - Dirac or Majorana -
of massive neutrinos by studying the process of RENP
is very challenging experimentally.
It  is discussed in detail in \cite{Dinh:2012qb}
and we will consider it very briefly here
for completeness. It is based on the fact that
the rate of emission of
a pair of Majorana neutrinos (particles) with masses
$m_i$ and $m_j$ in the threshold region
differs from the rate of emission (production) of a pair of
Dirac neutrinos (particles) with the same masses by
the presence of an interference term
$\propto m_im_j$ \cite{eechi1chi286} in
the emission (production) rate.
In the case under discussion
the interference term
is proportional to  $m_im_jB^{M}_{ij} =
m_im_j (1 - 2({\rm Im}(a_{ij}))^2/|a_{ij}|^2)$.
In the discussion which follows we neglect the effects
of the 4th neutrino which amounts to neglecting
corrections $\sim 10^{-2}$.
For $i=j$ we have $B^{M}_{ij} = 1$,
the interference term is negative and tends to suppress
the neutrino emission rate. In the case of $i \neq j$,
the factor $B^{M}_{ij}$, and thus the rate
of emission of a pair of different Majorana
neutrinos, depends on specific combinations
of the Majorana and Dirac CPV phases of the
neutrino mixing matrix, which in the case of the
reference 3-flavour neutrino mixing scheme were given
for the first time in  \cite{Dinh:2012qb}.
More specifically, negelcting corrections of the order of
$10^{-2}$, associated with the presence
of the 4th neutrino $\nu_4$ in the mixing
\footnote{It is not difficult to show that
the correction
 i) to  $B^{M}_{12}$ are of the order of
${\rm max}(s^2_{14}c_{12}s_{12},s_{14}s_{24}c^2_{12},s^2_{24}c_{12}s_{12})$,
ii) to $B^{M}_{13}$ are of the order of
${\rm max}(s^2_{14}c_{12}s_{13},s_{14}s_{24}c_{12}s_{23},s^2_{24}s_{12}s_{23})$,
and iii) to $B^{M}_{23}$ are of the order of
${\rm max}(s^2_{14}s_{13}s_{12},s_{14}s_{24}s_{12}s_{23},
s^2_{24}c_{12}c_{23}s_{23})$.
},

\begin{figure}[t]
\begin{center}
\begin{tabular}{cc}
\includegraphics[width=9.5cm,height=6.5cm]{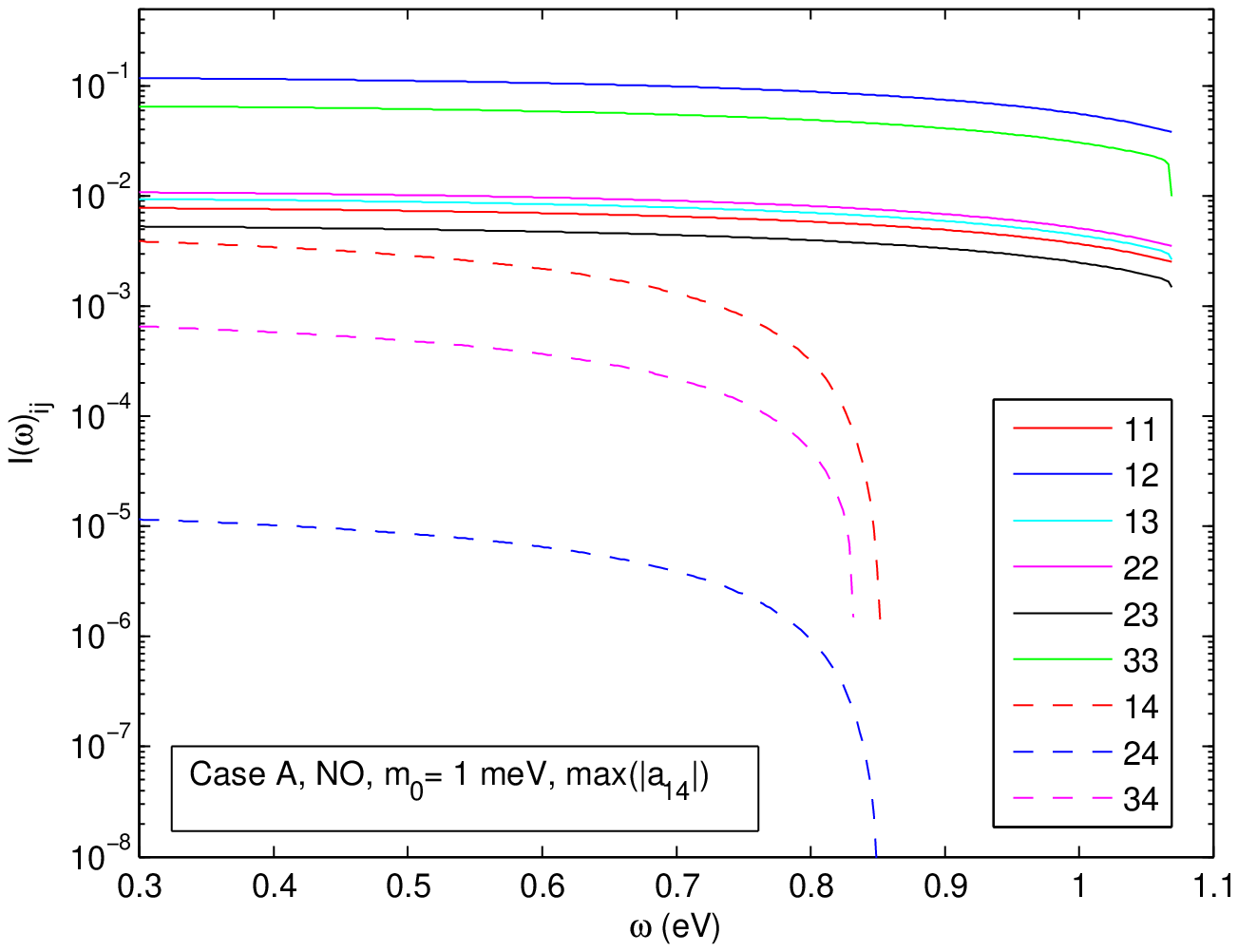}&
 \includegraphics[width=7.5cm,height=6.5cm]{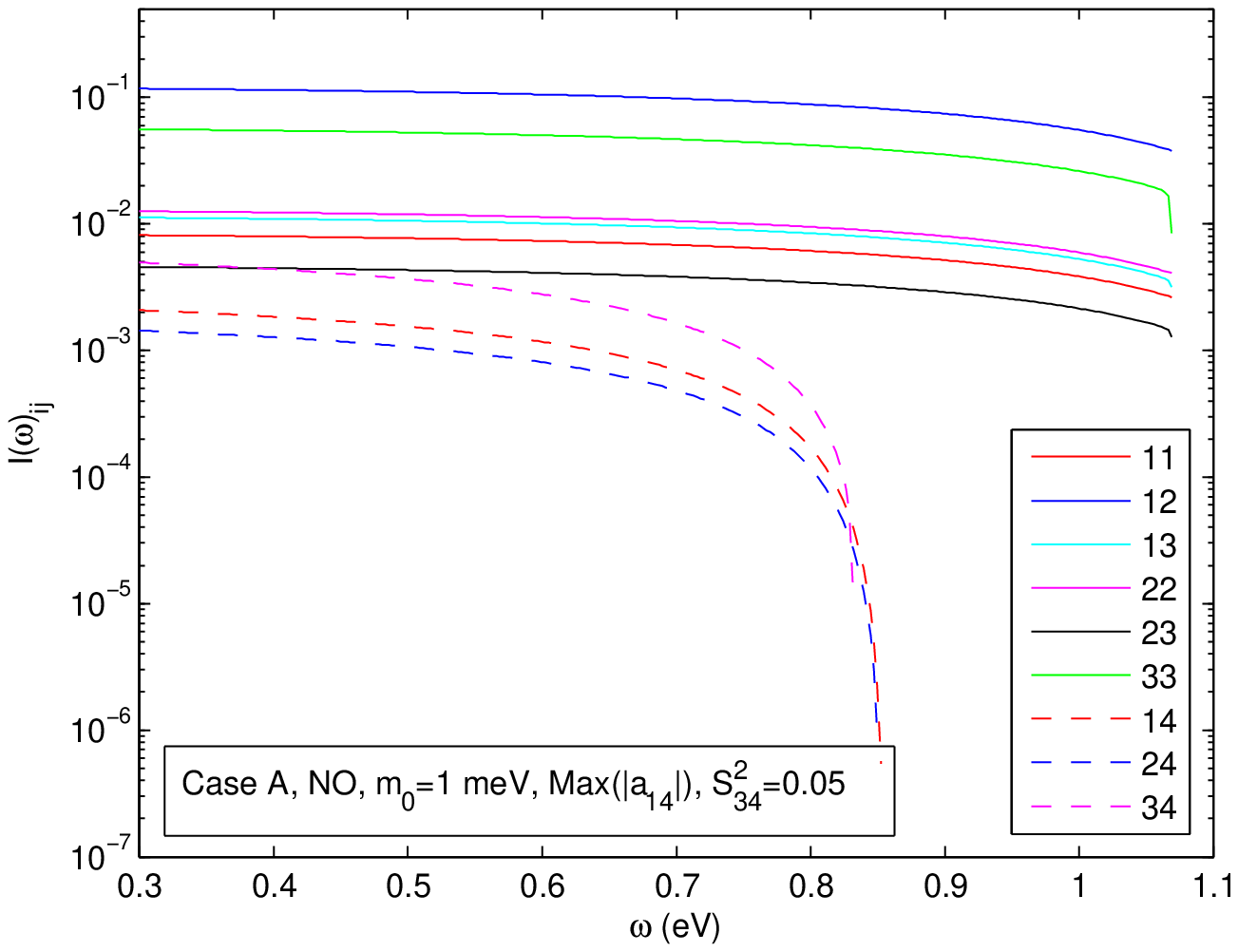}
\end{tabular}
\caption{The photon spectra corresponding
to the emission of the different individual neutrino pairs
 $(\nu_i + \nu_j)$, $I_{ij}(\omega)$, in the case of the NO
neutrino mass spetcrum with $m_0 = 1$ meV and
massive Dirac neutrinos. The spectra are obtained using
the set A of values of the oscillation parameters,
$\sin^2\theta_{34} = 0$ (left panel) and 0.05 (right panle),
and for the maximal value of  $|a_{14}|$.
The corresponding figures for IO neutrino mass spectrum
look practically the same.
}
\label{IijNOIO}
 \end{center}
\end{figure}
%

\noindent we find:
\beq
B^{M}_{12} \cong \cos \alpha^\prime_{21}\,,~
B^{M}_{13} \cong \cos (\alpha_{31} - 2\delta_{13})\,,~
B^{M}_{23} \cong \cos (\alpha^\prime_{21} - \alpha_{31} + 2\delta_{13})\,,~~
\alpha^\prime_{21}\equiv \alpha_{21} -2\delta_{21}\,.
\label{BMij}
\eeq
%
In contrast, the rate of emission of a pair
of Dirac neutrinos in the case of 3-neutrino
mixing does not depend on the CPV phases of the
PMNS matrix. If CP invariance holds we have
$\alpha_{21},\alpha_{31} = 0,\pi$, $\delta_{21} = 0,\pi$,
$\delta_{31} = 0,\pi$, and,
correspondingly, $B^{M}_{ij} = -1~{\rm or}~+1$, $i\neq j=1,2,3$.
For  $B^{M}_{ij} = +1$, the interference term
tends to suppress the neutrino emission rate, while for
$B^{M}_{ij} = -1$ it tends to increase it.
If, e.g., $\alpha_{21}$ has a CP violating value
we would have $-1 < B^{M}_{12} < +1$. Similar observation is valid
for $B^{M}_{13}$ and/or $B^{M}_{23}$ provided, e.g.,
 $\alpha_{31} \neq k\pi$ and/or
$\alpha_{21} - \alpha_{31}  \neq k^\prime\pi$,
$k,k^\prime=0,1,2$. Given the fact that,
as it follows from Table 1 in
\cite{Dinh:2012qb} as well as from Tables
\ref{tab_aijA} and \ref{tab_aijB},
we have $|a_{12}|^2 \gg |a_{13}|^2,|a_{23}|^2$,
the study of the emission  of the neutrino pair
$\nu_1 + \nu_2$ appears to be most promissing for
determination of the nature of massive neutrinos.
In the case of NH spectrum, however, the
term of interest  $m_1m_2B^{M}_{12}$
can be strongly suppressed due to the
relatively small value of $m_1$.
No such a suppression can take place
for the IO spectrum, including the
IH case. Futher details regarding the problem
of determination of the nature
of the light massive neutrinos
$\nu_{1,2,3}$ by measuring the spectrum of the
photon emitted in the process of RENP
can be found in \cite{Dinh:2012qb}.

%
%
\subsection{The RENP Phenomenology of
Emission of the 4th Neutrino
with Mass at the eV Scale
}
%

 It follows from Tables \ref{tab_aijA}, \ref{tab_aijB} and
\ref{tab_aijA2}
that, in what concerns the heaviest 4th neutrino $\nu_4$,
for $\sin^2\theta_{34} = 0~(0.05)$
the largest maximal value $\sim (0.09 - 0.10)$
($\sim 0.14$) have the factors $|a_{14}|$ and $|a_{24}|$
(has the factor  $|a_{34}|$), while the largest minimal
value $\sim 0.04$~($\sim 0.12$)  in the cases A and B have
respectively the factors
$|a_{34}|$ and $|a_{14}|$). Given the fact that
the largest $|a_{ij}|$ factor is $|a_{12}| \sim (0.44 - 0.45)$
and corresponds to the emission of the $(\nu_1 + \nu_2)$ pair,
the emission of the  heaviest 4th neutrino $\nu_4$,
even sufficiently far from the threshold,
will proceed with rate which is for
$\sin^2\theta_{34} = 0~(0.05)$
at least by a factor $\sim 20$ ($\sim 10$)
smaller than the rate of
emission of the $(\nu_1 + \nu_2)$ pair.
Near the threshold it will be further suppressed.

 The predicted rate of emission of each of the individual pairs
 $(\nu_i + \nu_j)$, $I_{ij}(\omega)$,
in the case of the NO  spetcrum with $m_0 = 1$ meV,
for the set A of values of the neutrino oscillation parameters
with $\sin^2\theta_{34} = 0$ (left panel) and
$\sin^2\theta_{34} = 0.05$ (right panel)
and maximal  $|a_{14}|$, is shown in Fig. \ref{IijNOIO}
as a function of the photon energy $\omega$~
\footnote{We do not show the corresponding figures for
IO neutrino mass spectrum beacause they are very similar
to those shown for NO spectrum.
}.
Increasing $m_0$ up to $m_0 = 100$ meV leads
to practically the same results for
$I_{ij}(\omega)$ at $\omega$ sufficiently smaller
than $\omega_{ij}$.
As Figs. \ref{IijNOIO} suggests and the preceding
considerations imply, observing the contribution
to the photon spectral rate
due to the emission of pairs of neutrinos
at least one of which is $\nu_4$
would require a relatively high precision
measurement of the photon spectrum
at $\omega < \omega_{14}$.
The same conclusion is valid for
the set B of values of the neutrino
oscillation parameters ($\sin^2\theta_{34} = 0$).
\begin{figure}[t]
\begin{center}
\begin{tabular}{cc}
\includegraphics[width=9.5cm,height=6.5cm]{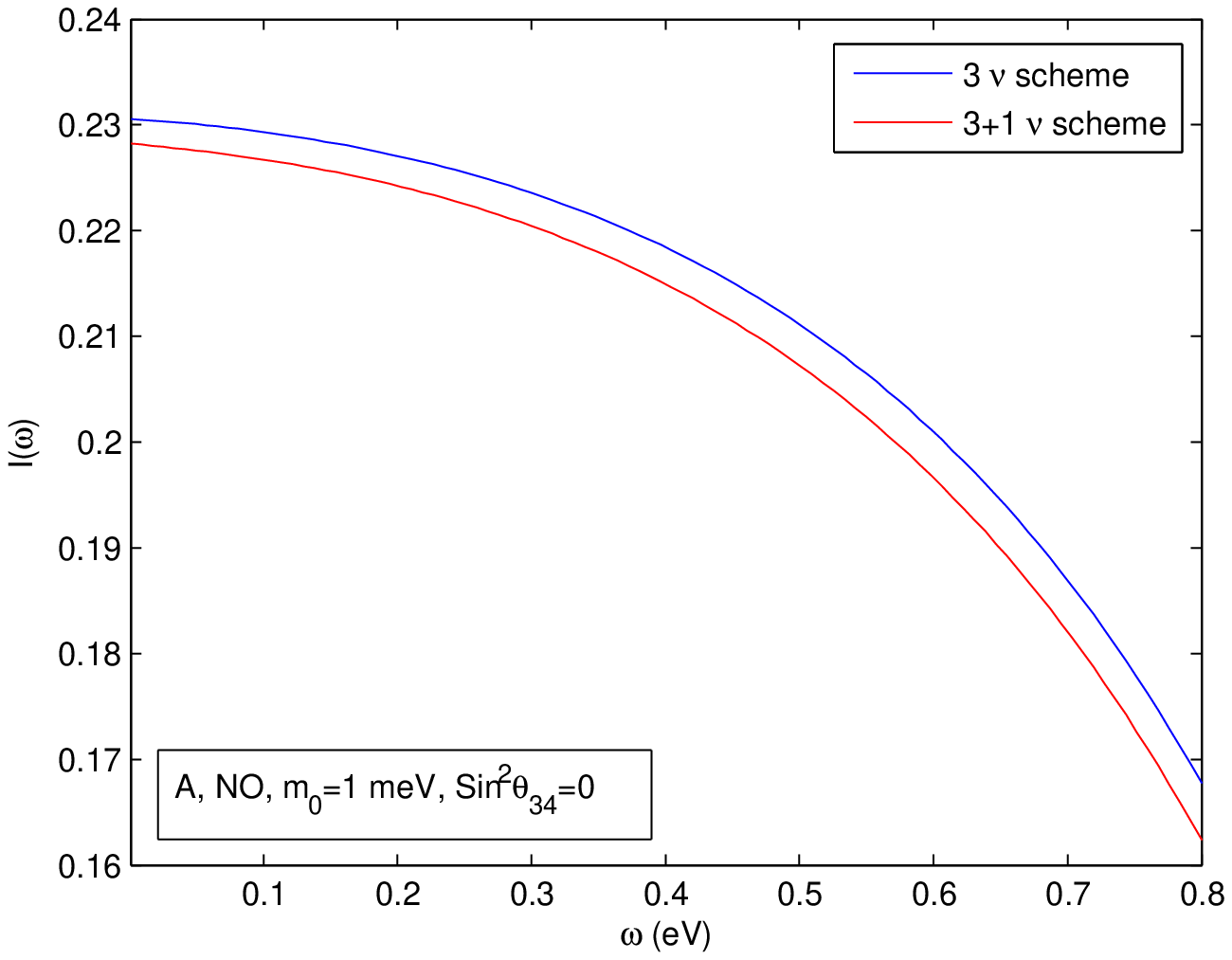}&
\includegraphics[width=7.5cm,height=6.5cm]{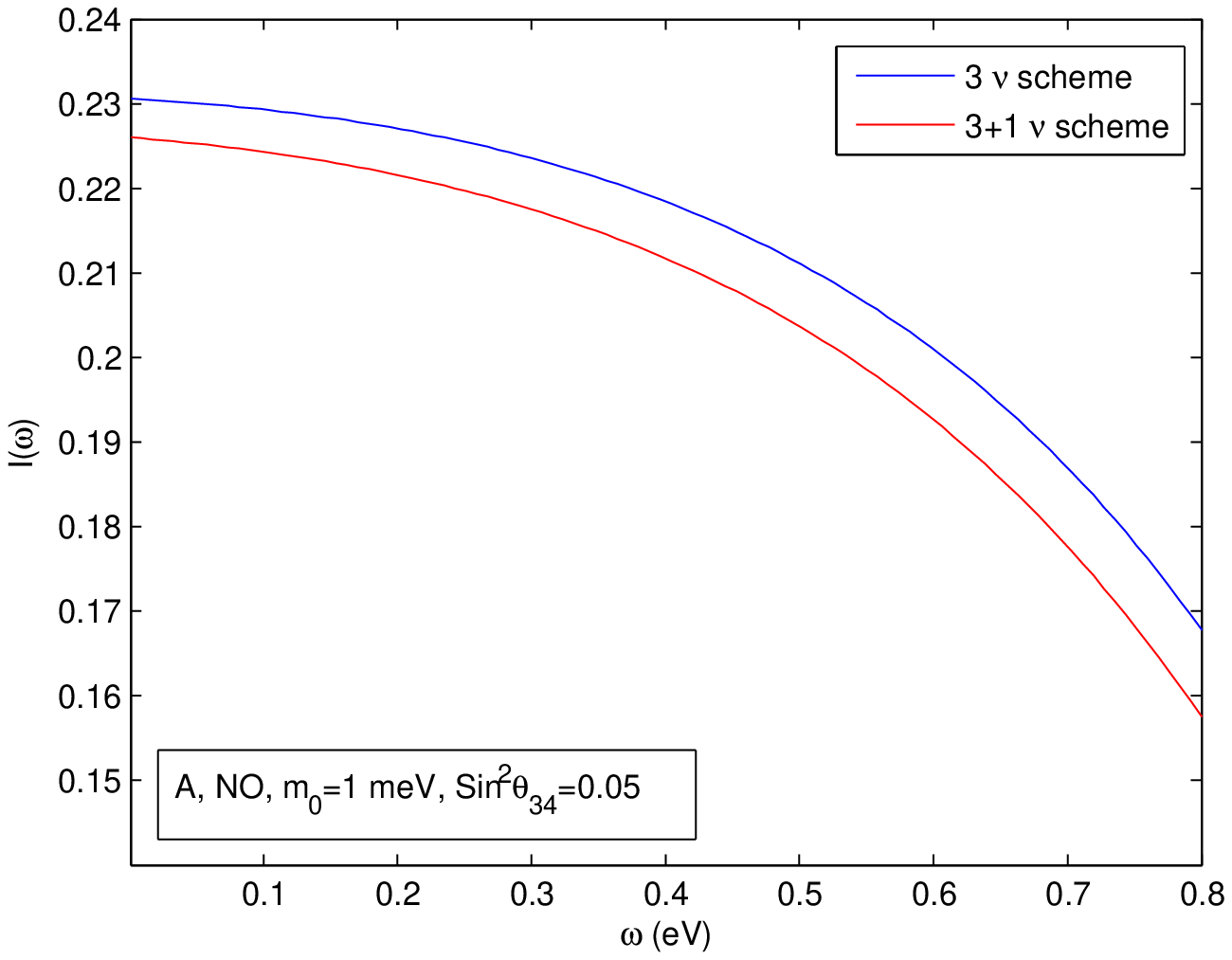}
\end{tabular}
\caption{The spectral rates of $3$-neutrino and $(3+1)$-neutrino schemes for the set A of neutrino oscillation data with
$\sin^2\theta_{34}=0$ (left panel) and $\sin^2\theta_{34}=0.05$ (right panel).
}
\label{3nuvs4nu}
 \end{center}
\end{figure}
%
 In Fig. \ref{3nuvs4nu} we present
for NO neutrino mass spectrum
the sum $I^{3\nu}(\omega) =
\sum_{i,j=1,2,3}I^{3\nu}_{ij}(\omega)$,
and the sum   $I^{(3+1)} = \sum_{i,j=1,2,3,4}I_{ij}(\omega)$,
as functions of $\omega$ in the region
$\omega < \omega_{34}$, i.e.,
below the threshold of production of the
neutrino pair $\nu_3 + \nu_4$
with the largest sum of masses
\footnote{In the case of the
transitions between the atomic energy levels
considered and the values of $\Delta m^2_{14(34)}$
used in our analysis,
the pair $\nu_4 + \nu_4$
cannot be emitted in the
process of interest.
}.
The spectral rates $I^{3\nu}_{ij}(\omega)$,
$i,j=1,2,3$, have been computed assuming
3-neutrino mixing, i.e., no presence of sterile
neutrinos in the mixing, while
the rates  $I_{ij}(\omega)$
have been calculated in the $3+1$ scheme
for the case A of sterile neutrino
oscillation parameters and
$\sin^2\theta_{34} = 0$ and 0.05.
As Fig. \ref{3nuvs4nu} shows,
the total spetral rates
$I^{3\nu}(\omega)$ and
$I^{(3+1)}(\omega)$
differ by approximately
(0.005 - 0.010) for $\omega$ sufficiently
smaller than the threshold
energy $\omega_{34}$.
This differnce is independent of the
the value of $m_0\leq 100$ meV.
The same conclusion is valid in the case of the
IO neutrino mass spectrum.
The indicated difference can be used,
in principle, to test
the hypothesis of existence
of a 4th (sterile) neutrino with
a mass at the eV scale
in a RENP type of experiment.
Such a test would require
a rather precise calculation
of the the total spectral rate
$I^{3\nu}(\omega)$
at sufficiently small values of
$\omega < \omega_{ij}$,
$i,j=1,2,3$. Given the fact that
at the values of $\omega$
of interest $I^{3\nu}(\omega)$
is practically independent
of the values of the neutrino masses,
a calculation of $I^{3\nu}(\omega)$
with the requisite precision
might not be impossible.

  Further, the threshold energy $\omega_{14}$ for the
emission of the $\nu_1 + \nu_4$ pair
is well separated from the energy thresholds
of the emission of pairs
of the 3 light neutrinos  $\omega_{ij}$,
$i,j=1,2,3$. This is clearly seen also
in Fig. \ref{IijNOIO}.
Indeed, for the NO (IO) spectrum
with $m_0= 0.001;~0.01$ eV, for instance,
the thresholds $\omega_{ij}$ for $i,j=1,2,3$
are grouped in the vicinity of 1.07 eV,
while  $\omega_{14} \cong 0.85$ eV
($\omega_{14} \cong 0.83$ eV).
Similar results are valid for $m_0 = 0.10$ eV.

The threshold  $\omega_{34}$ is relatively close to
$\omega_{14}$. Indeed, for, e.g.,
$m_0 = 0.001~(0.100)$ eV and NO neutrino mass
spectrum we have $\omega_{14} = 0.8544~(0.8049)$
and $\omega_{34} = 0.8319~(0.7991)$.
For IO spectrum and the same values of
$m_0$ we get $\omega_{14} = 0.8325~(0.7993)$
and $\omega_{34} = 0.8544~(0.8049)$.
Distinguishing between the threshold energies
 $\omega_{14}$  and $\omega_{34}$ would not be a problem,
in principle, since it is expected that
in the RENP experiments
the photon energy will be known with a
relative uncertainty of $\sim 10^{-5}$.
It should be added, however, that for the values of
$\delta_{12} = \delta_{13} = 0$, for which
$|a_{14}|^2$ has a maximal value, and , e.g.,
for the set A of the neutrino oscillation parameters,
$|a_{34}|^2$ is approximately by a factor of 6
smaller than $|a_{14}|^2$.

 In the case of Majorana neutrinos,
the factor $B^{M}_{14}$, associated
with the emission of $\nu_1 + \nu_4$,
is given by a somewhat lengthy  expression.
For $\delta_{12},\delta_{13} = 0$ or $\pi$ we have:
\beq
B^{M}_{14} = \cos \alpha_{41}\,.
\label{BM14}
\eeq
%

\begin{figure}[t]
\begin{center}
\begin{tabular}{cc}
\includegraphics[width=9.5cm,height=6.5cm]{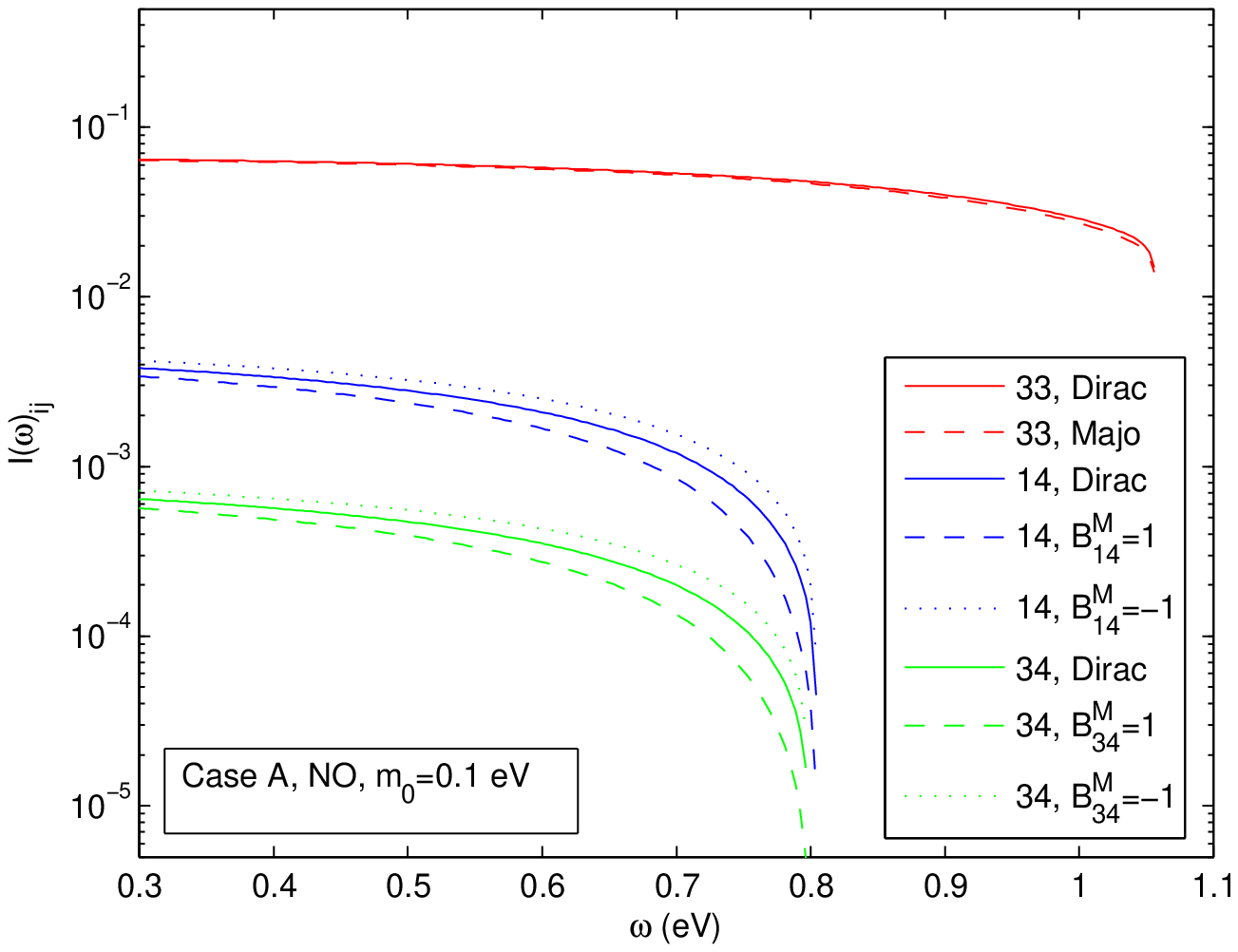} &
\includegraphics[width=7.5cm,height=6.5cm]{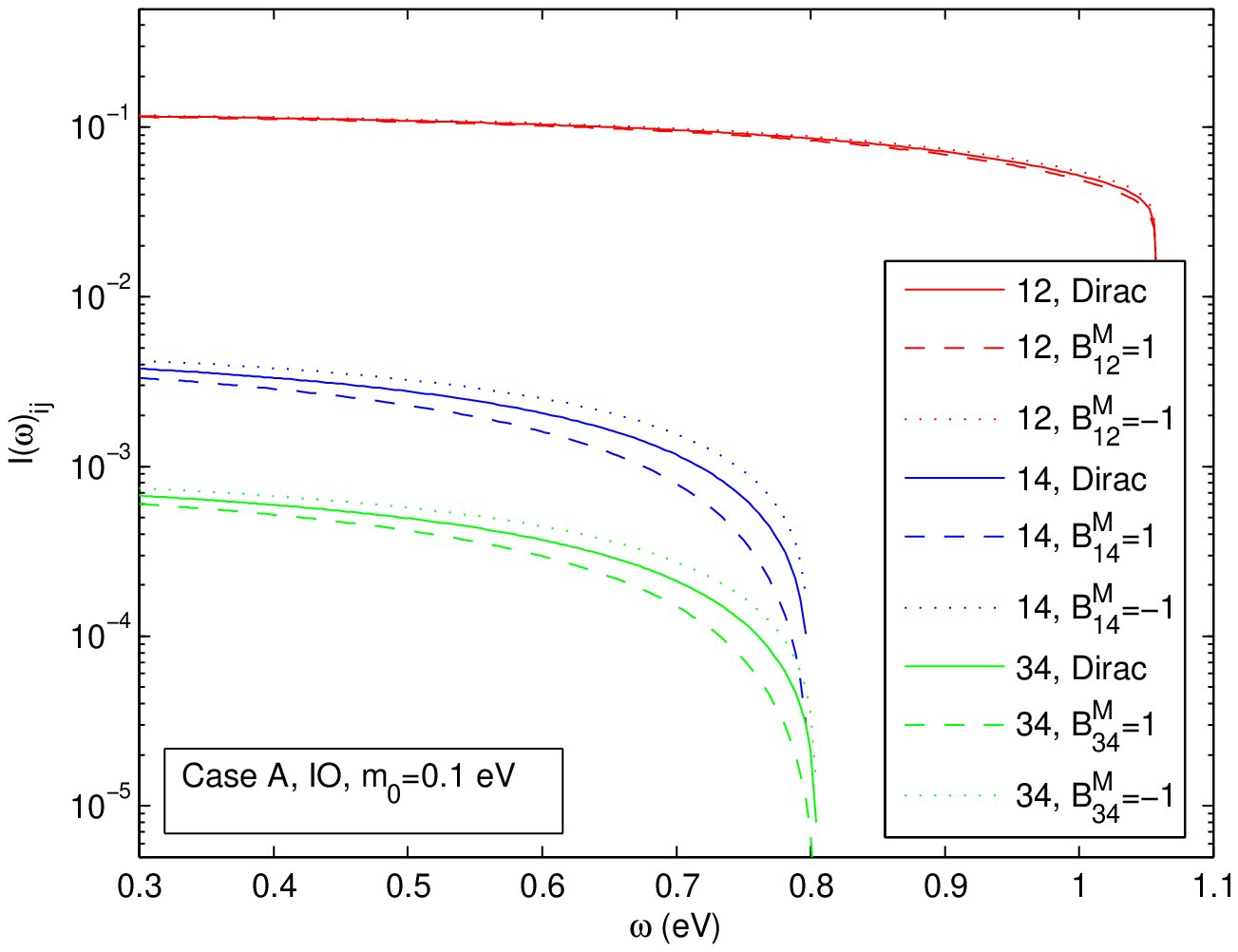}
\end{tabular}
\caption{The photon spectra
 $I_{14}(\omega)$ and  $I_{34}(\omega)$
in the case of i) Dirac neutrinos (solid lines),
ii) Majorana neutrinos and $B^{M}_{14},B^{M}_{34} = +1$
(dashed lines), and iii) Majorana neutrinos and
$B^{M}_{14},B^{M}_{34} = -1$ (dotted lines),
in the case of the NO (left panel) and IO (right panel)
neutrino mass spetcra with $m_0 = 0.1$ eV.
We show for comparison also the spectrum  $I_{33}(\omega)$
($I_{12}(\omega)$) in the left (right) panel.
The spectra are obtained using
the set A of values of the oscillation
parameters and for $\delta_{12} = \delta_{13} = 0$
(i.e., for maximal $|a_{14}|$), $\alpha_{31} = 0$
 and $\alpha_{21} = \pi~(0.517)$ (left panel)
 and
$\alpha_{21} = 0~(\pi)$ (right panel),
corresponding, respectively, to $B^{M}_{34} = +1~(-1)$
and  $B^{M}_{12} = +1~(-1)$.
}
\label{I1434NOIO}
 \end{center}
\end{figure}
%

 In Fig. \ref{I1434NOIO} we illustrate the effects
of the term  $\propto m_im_jB^{M}_{ij}$ on the individual spectra
$I_{14}(\omega)$ and  $I_{34}(\omega)$ of emission
of the pairs of Majorana neutrinos $\nu_1 + \nu_4$ and
$\nu_3 + \nu_4$ in the cases of NO (left panel) and IO (right panel)
neutrino mass spectra with $m_0 = 0.1$ eV.
We show for comparison also the spectrum  $I_{33}(\omega)$
($I_{12}(\omega)$)  in the left (right) panel
at $\omega <\omega_{33}$
($\omega <\omega_{12}$) outside the
thershold region.
The spectra are obtained using
the set A of values of the oscillation
parameters and for $\delta_{12} = \delta_{13} = 0$
(i.e., for maximal $|a_{14}|$), $\alpha_{31} = 0$
and  $\alpha_{21} = \pi~(0.517)$ (left panel) and
$\alpha_{21} = 0~(\pi)$ (right panel),
corresponding, respectively, to $B^{M}_{34} = +1~(-1)$
and  $B^{M}_{12} = +1~(-1)$.
As Fig. \ref{I1434NOIO} indicates, the effect of the
terms $\propto m_im_jB^{M}_{ij}$ on the spectra
$I_{14}(\omega)$ and  $I_{34}(\omega)$
can be sizable in the threshold region.
In the case of NO spectrum, the term
$\propto m_1m_4B^{M}_{14}$
can be suppressed due to a small value of the
lightest neutrino mass $m_1$.
Such a suppression will not hold for IO spectrum.
However, the observation of the effects of
the terms $\propto m_1m_4B^{M}_{14}$ and/or
$\propto m_3m_4B^{M}_{34}$  on the total
photon spectrum $I(\omega)$ in the case of massive Majorana
neutrinos is very challenging due to the
relatively small values of the factors
 $|a_{14}|^2$ and $|a_{34}|^2$. Such small values
are a consequence of the relatively small
phenomenologically allowed couplings of the
the 4th (sterile) neutrino $\nu_4$  to the
electron in the weak charged lepton current.

The results described in the present Section
have been obtained for the best fit values of the
sterile neutrino oscillation parameters
quited in eqs. (\ref{A1}) and (\ref{B1}).
In ref. \cite{Giunti:2013vaa} the $3\sigma$
allowed ranges of the parameters of interest were
also reported. For $\Delta m^2_{41(43)}$
this range reads:
$0.82~{\rm eV^2} \ltap \Delta m^2_{41(43)} \ltap 2.19~{\rm eV^2}$.
The $3\sigma$ maximal values of $\sin^2\theta_{14}$ and
$\sin^2\theta_{24}$ found in \cite{Giunti:2013vaa} are the following:
$\sin^2\theta_{14}\ltap 0.058$ and $\sin^2\theta_{24}\ltap 0.016$.
For $\sin^2\theta_{14} = 0.058$ and $\sin^2\theta_{24} = 0.016$,
the values of the factors $|a_{ij}|$
are given in Table 3. Comparing them with the values quoted in Tables 1 and 2
we see that the RENP rate for emission of the pair of neutrinos
$(\nu_1 + \nu_4)$ can be approximately by a factor 1.8
larger than the rate predicted using the best fit values of
$\sin^2\theta_{14}$ and $\sin^2\theta_{24}$
quoted in eqs. (\ref{A1}) and (\ref{B1}).
The rates of emission of the neutrino pairs $(\nu_2 + \nu_4)$ and
$(\nu_3 + \nu_4)$ can also be larger, but by smaller factors.
In the case of the atomic levels considered the emission of the
pair $(\nu_4 + \nu_4)$ cannot take place
because the energy of the transition available for the
emission of the neutrinos is smaller than 2$m_4$.
\begin{table}
\centering \caption{
\label{tab_aijB2}
The quantity $|a_{ij}|=|U^*_{ei}U_{ej}-
\frac{1}{2}\sum_{l=e,\mu,\tau}  U_{l i}^*U_{l j}|$ (NO),
 case B, $\sin^2\theta_{14} = 0.058$, $\sin^2\theta_{24} = 0.016$.
}
    \begin{tabular}{|c|c|c|c|c|}
        \hline
        $|a_{11}|$      & $|a_{12}|$      & $|a_{13}|$      & $|a_{14}|$      & $|a_{22}|$      \\ \hline
        $0.1410-0.1679$ & $0.4273-0.4396$      & $0.1113-0.1328$      & $0.0748-0.1354$ & $0.1937-0.2181$ \\ \hline
        $|a_{23}|$      & $|a_{24}|$      & $|a_{33}|$      & $|a_{34}|$      & $|a_{44}|$      \\  \hline
        $0.0733-0.0912$      & $0.0289-0.1113$ & $0.4707-0.4773$ & $0.0200-0.0591$ & $0.0265$ \\ \hline
    \end{tabular}
   \end{table}
%
%
\section{Summary and Conclusions}

We have analysed the possibility to test
the hypothesis of exsitence of
neutrinos with masses at the eV scale
coupled to the electron in the weak charged
lepton current in an atomic physics experiment
on radiative emission of neutrino pair
(RENP), in which  the spectrum of the photon
is measured with high precision.
The RENP is a process of  collective de-excitation of
atoms in a metastable level into emission
mode of a single photon plus a neutrino pair \cite{RENP1}.
The process of RENP was shown to be sensitive
to the absolute values of the masses of the
emitted neutrinos, to the type of spectrum
the neutrino masses obey and
to the nature - Dirac or Majorana - of
massive neutrinos \cite{RENP1,Dinh:2012qb}.
If more than three light neutrinos
couple to the electron in the weak charged
lepton current and the additional neutrinos
beyond the three known have masses at the
eV scale, they will be emitted
in the RENP process. This will lead
to new observable features in the spectrum of the
photon, emitted together with the neutrino pair.
The presence of eV scale neutrinos in the
neutrino mixing is associated with the
existence of sterile neutrinos
which mix with the active flavour
neutrinos. At present there are a number
of hints for active-sterile neutrino
oscillations driven by
$\Delta m^2 \sim 1~{\rm eV^2}$.
In the present article we have investigated these features,
concentrating for simplicity on the $3 + 1$
phenomenological model with one sterile neutrino.
We have used two sets of values of the
three additional neutrino mixing parameters of
the $3 + 1$ model relevant for our study -
$\sin^2\theta_{14}$, $\sin^2\theta_{24}$ and  $\Delta m^2_{41(43)}$ -
given in  eqs. (\ref{A1}) and (\ref{B1}).
These values were found in  the global analyses of all
the data (positive evidences and negative results)
relevant for the tests of the sterile neutrino hypothesis,
performed in \cite{Kopp:2013vaa} and
 \cite{Giunti:2013vaa}, respectively.

The emission of the neutrino pair $(\nu_i + \nu_j)$
will lead to a kink in the photon spectrum
at the threshold energy $\omega_{ij} = \omega_{ji}$.
For three massive neutrinos of the ``standard'' 3-neutrino mixing scheme
there are altogether 6 different pairs $\nu_1+\nu_1$, $\nu_1+\nu_2$,...,
 $\nu_3+\nu_3$, and, correspondingly,
6 threshold energies $\omega_{ij}$.
In the ``3 + 1'' model there are 4 additional pairs
$\nu_1 + \nu_4$, $\nu_2+\nu_4$, $\nu_3+\nu_4$ and $\nu_4+\nu_4$,
and therefore altogether 10 thresholds.
The magnitude of the contribution of the $(\nu_i + \nu_j)$
emission to the photon spectral rate
is determined essentially by the factor
$|a_{ij}|^2$ (eq. (\ref{aij})), which depends only on
the neutrino mixing angles and the CP violation phases present
in the neutrino mixing matrix.
For the two sets of the
sterile neutrino oscillation parameters
considered by us the possible values of
the factors $|a_{ij}|$ are given in Tables 1 and 2.

 We have shown, in particular, that
the presence of the 4th neutrino of the ``3+1'' scheme
leads to an overall decreasing of the photon  spectral rate
near the prominent 3-neutrino mixing threshold
 $\omega_{33}$  ($\omega_{12}$)
by approximately 0.002 (0.001)  with respect to the rate
corresponding to the 3-neutrino mixing.
The quoted magnitude of the change of the spectral rate
remains practically the same for values of
 the lightest neutrino mass
$m_0 \ltap 0.10$ eV.
The observation of the indicated relatively
small difference between the two photon
spectral rates under discussion,
corresponding  the 3-neutrino and
the (3+1)-neutrino mixing,
requires a high precision measurement
of the photon spectrum.
The quoted result illustrates the more
general conclusion reached in the present study, namely, that
the presence of a 4th (sterile) neutrino
in the mixing in the ``3 + 1'' scheme
has a little effect on the sensitivities of the RENP process
to the masses and the mixing of the three lighter neutrinos
$\nu_{1,2,3}$, i.e., the absolute neutrino
mass scale, the type of the neutrino
mass spectrum and the nature of massive neutrinos,
associated with the sub-mixing of
the 3 active neutrinos.

 The threshold energy $\omega_{14}$ for the
emission of the $\nu_1 + \nu_4$  ($\nu_3 + \nu_4$)
pair which in the cases A with
$\sin^2\theta_{34} = 0$ ($\sin^2\theta_{34} = 0.05$)
has the largest $|a_{ij}|$
factor for $i=1,2,3,4$ and $j=4$
(i.e., for emission of a pair of neutrinos at least one of which is
the heaviest neutrino $\nu_4$),
was shown to be  well separated from the energy thresholds
of the emission of pairs
of the 3 light neutrinos  $\omega_{ij}$,
$i,j=1,2,3$.
As the numerical analysis
performed by us allowed to conclude,
the emission of the  heaviest 4th neutrino $\nu_4$,
even sufficiently far from the threshold,
is predicted to proceed with rate which is at least by a
factor $\sim 20$ ($\sim 10$)
smaller than the rate of
emission of the lighter $(\nu_1 + \nu_2)$ pair
having the largest $|a_{ij}|$ factor.
Near the threshold it will be futher
suppressed.
 At values of the photon energy
$\omega$, which are sufficiently smaller than
${\rm min}(\omega_{ij})$, $i,j=1,2,3,4$,
the total spectral rate
in the case of mixing of 3 neutrinos only,
i.e., no presence of sterile neutrino(s)
in the mixing,
$I^{3\nu}(\omega) =\sum_{i,j=1,2,3}I^{3\nu}_{ij}(\omega)$,
and the total spectral rate
in the $3+1$ scheme,
$I^{(3+1)}(\omega) = \sum_{i,j=1,2,3,4}I_{ij}(\omega)$,
differ approximately by 0.010.
This difference can be used,
in principle, to test
the hypothesis of existence
of a 4th (sterile) neutrino with
a mass at the eV scale
in a RENP type of experiment.
Such a test would require
a rather precise calculation
of the the total spectral rate
$I^{3\nu}(\omega)$
at sufficiently small values of
$\omega < \omega_{ij}$,
$i,j=1,2,3$. At the values of
$\omega$ of interest
the 3-neutrino total
spectral rate $I^{3\nu}(\omega)$
is practically independent
of the neutrino masses
and a calculation of
$I^{3\nu}(\omega)$ with the requisite
precision might not be impossible.

 The results obtained in the present study
show that observing in an RENP experiment
the contribution to the photon spectral rate
due to the emission of pairs of neutrinos
at least one of which is the eV scale neutrino $\nu_4$
of the ``3+1'' scheme with one sterile neutrino
would be very challenging since it would
require a relatively high precision
measurement of the photon spectral rate.

\section*{Acknowledgments}
This work was supported in part
by the Vietnam National Foundation for
Science and Technology Development (NAFOSTED)
under the grant 103.03-2012.49 (D.N.D.),
by the INFN program on
``Theoretical Astroparticle Physics'' (TASP),
by the research grant  2012CPPYP7 ({\it  Theoretical Astroparticle Physics})
under the program  PRIN 2012 funded by the Italian Ministry of Education,
University and Research (MIUR),
by the World Premier International Research Center
Initiative (WPI Initiative), MEXT, Japan,
and by the European Union FP7-ITN INVISIBLES and UNILHC
(Marie Curie Action, PITAN-GA-2011-289442 and PITN-GA-2009-23792)
(S.T.P.).

\end{document}